\RequirePackage{fix-cm}
\documentclass[natbib]{svjour3}                     
\smartqed  
\usepackage{graphicx}
\usepackage{epsf}
\usepackage{amssymb}
\usepackage{amsmath}
\usepackage{placeins}
\usepackage{color}

 \journalname{SSRv}
\newcommand{\be}{\begin{equation}}
\newcommand{\ee}{\end{equation}}
\newcommand{\beq}{\begin{eqnarray}}
\newcommand{\eeq}{\end{eqnarray}}
\newcommand\subsun[1]{{$_{\normalsize\odot}$}}

\newcommand{\ls}{{_<\atop^{\sim}}}

\begin{document}
\title{The Morphologies and Kinematics of Supernova Remnants}

\titlerunning{Morphology and Kinematics of SNRs}        

\author{Laura~A.~Lopez \and Robert~A.~Fesen
}

\authorrunning{Lopez \& Fesen} 

\institute{Laura~A.~Lopez \at
The Ohio State University, 140 W. 18th Ave., Columbus, OH 43210, USA \\
\email{lopez.513@osu.edu}
\and
Robert~A.~Fesen \at
              Dartmouth College, 6127 Wilder Lab, Hanover, NH 03755, USA \\
              \email{robert.a.fesen@dartmouth.edu}
                              }

\date{Received: 15 January 2018 / Accepted: 26 January 2018}

\maketitle

\abstract{We review the major advances in understanding the morphologies and kinematics of supernova remnants (SNRs). Simulations of SN explosions have improved dramatically over the last few years, and SNRs can be used to test models through comparison of predictions with SNRs' observed large-scale compositional and morphological properties as well as the three-dimensional kinematics of ejecta material. In particular, Cassiopeia A -- the youngest known core-collapse SNR in the Milky Way -- offers an up-close view of the complexity of these explosive events that cannot be resolved in distant, extragalactic sources. We summarize the progress in tying SNRs to their progenitors' explosions through imaging and spectroscopic observations, and we discuss exciting future prospects for SNR studies, such as X-ray microcalorimeters.}

\section{Introduction}

The predictive power of supernova explosion (SN) simulations has improved dramatically in the last few years. Three-dimensional simulations that properly handle energy-dependent neutrino transport obtain successful explosions via neutrino heating \citep{lentz15,janka16,roberts16,muller17}. These models are beginning to make testable predictions, such as the expected light curves for different mass progenitors, neutron star (NS) kick velocities, and large-scale compositional asymmetries (e.g., \citealt{wong13,utrobin17,wong17}). 

Although hundreds of SNe are discovered each year by dedicated surveys (e.g., \citealt{law09,leaman11,holoien17}), they are often too distant to resolve the SN ejecta and the immediate surroundings of the exploded stars (e.g., $1^{\prime \prime}\approx$ 50~pc for a distance of 10~Mpc). While study of the closest SNe, such as SN~1987A \citep{mccray16}, has advanced the field tremendously, our understanding of SN progenitors and the details of the explosion dynamics is severely hampered by the infrequency of nearby events.

Young, nearby supernova remnants (SNRs) offer the means to study SN explosions and dynamics at sub-pc scales, and thus they are crucial to test the predictions of SN models. Evolved SNRs are observable for up to 10$^{5}$~years after the explosions across the electromagnetic spectrum, and multiwavelength campaigns have identified over 500 SNRs in the Milky Way and nearby galaxies \citep{badenes10,sasaki12,green14,maggi16,garofali17}. Detailed investigations of SNRs that are a few thousand years old or less have yielded valuable insights into the progenitors, environments, and explosion dynamics for all SN subtypes (see e.g., \citealt{vink12} for a review). 

Cassiopeia~A (Cas~A; d $\simeq$ 3.4 kpc) is perhaps the most well-studied SNR in the Milky Way as it is the youngest known ($\sim$350~years old: \citealt{thorstensen01,fesen06b}) core-collapse SNR in the galaxy. Cas~A resulted from an asymmetric Type IIb explosion \citep{krause08,rest11} associated with a red supergiant progenitor with an initial mass of 15--25 $M_{\odot}$ \citep{young06}. The {\it Chandra} X-ray Observatory first-light image of Cas~A revealed the neutron star produced in the SN \citep{pavlov00}, and further X-ray investigations showed metal-rich ejecta primarily distributed in a $\approx$4$^{\prime}$ shell (e.g., \citealt{hughes00,hwang00,willingale02}). Kinematic studies of Cas~A have mapped the three-dimensional structure of the ejecta (e.g., \citealt{delaney10,gref17}), including the primary ring of SN debris as well as the high-velocity knots (up to 15,000 km s$^{-1}$) that protrude beyond the shell in the northeast and southwest \citep{fesen96,fesen01,fesen06b,fesen16}. 

The morphologies and kinematics of young SNRs like Cas~A are important observables to test and constrain the SN explosion models being developed now. In this Chapter, we summarize the recent advances in studies of SNR morphologies ($\S$\ref{sec:morphologies}) and SNR kinematics ($\S$\ref{sec:kinematics}), particularly related to Cas~A. We discuss how these results can be utilized to investigate the explosive origins of SNRs, the environments of SNRs, and the instabilities that mix ejecta with the surrounding medium.

\section{Morphologies of SNRs} \label{sec:morphologies}

\begin{figure}[h!]
\begin{center}
\includegraphics[width=\textwidth]{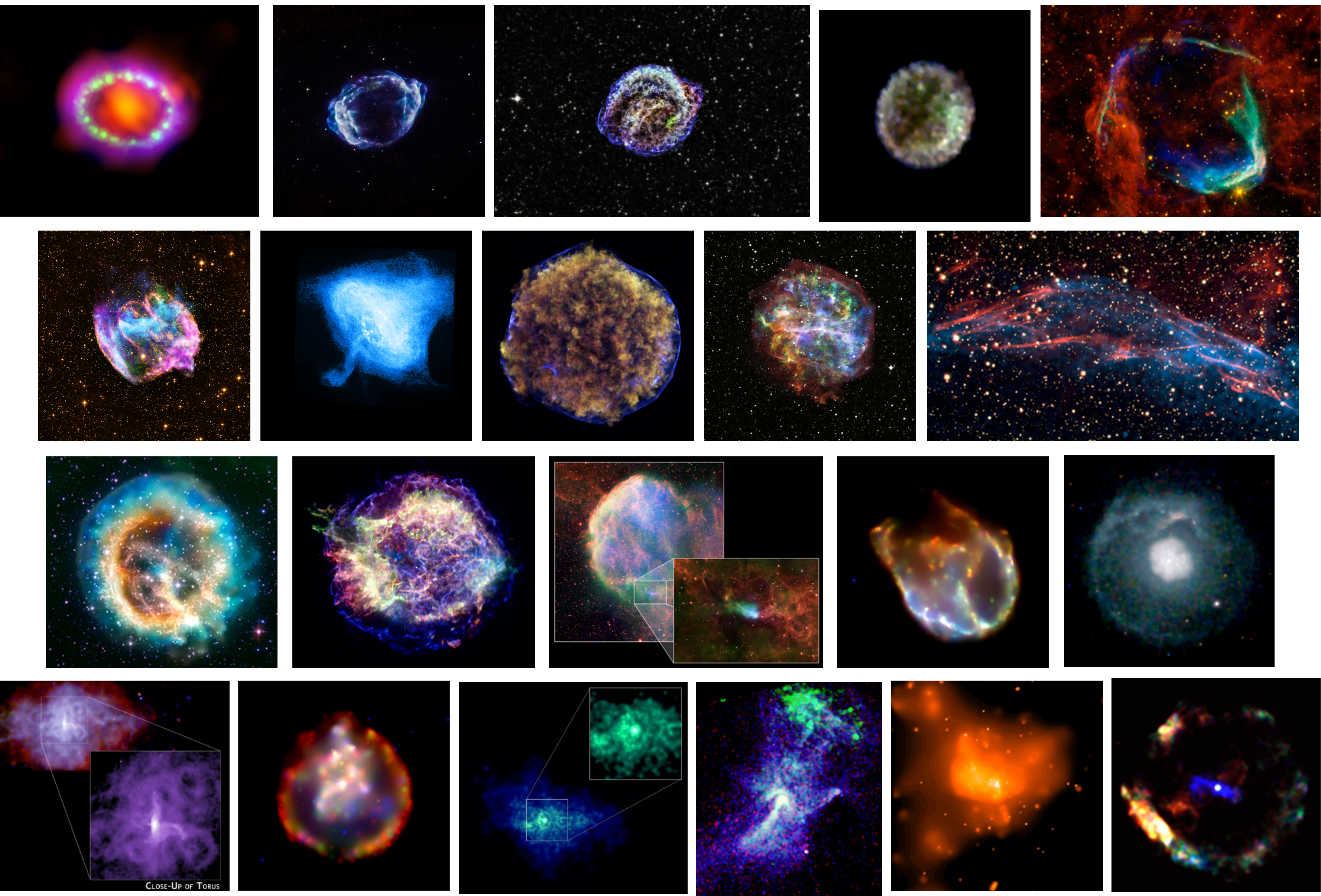} 
\end{center}
\caption{{\it Chandra} X-ray Observatory images of a wide array of SNRs. Their morphologies are complex and heterogeneous, yet they are a rich sample from which to learn about explosions and their environments. These images were adapted from the {\it Chandra} press webpage (http://www.chandra.harvard.edu/press/). \label{fig:SNRchandra}}
\end{figure}

The morphologies of SNRs are complex and varied (see Figure~\ref{fig:SNRchandra}). Consequently, significant effort has been made in terms of observations and interpretation of individual sources. However, through the decades, as more SNRs were imaged at radio, optical, and X-ray wavelengths, astronomers began to classify the sources based on their morphological characteristics \citep{mathewson83,seward90,williams99,green14}. 

Remnant classifications used most commonly in the field (e.g., as in the Green Catalog of Galactic SNRs: \citealt{green14}) are shell-type, center-filled, and composite. Shell-type SNRs are those whose emission is dominated by a limb-brightened shell (e.g., SN~1006: \citealt{winkler14}). Center-filled SNRs have pulsar wind nebulae (PWNe; see review by \citealt{gaensler06}) that shine brighter than their shells (e.g., the Crab: \citealt{seward06}). Composite SNRs are sources that have evidence of both a PWN and a shell in their radio and X-ray morphologies (e.g., MSH~11$-$56: \citealt{temim13}). From X-ray imaging with {\it Einstein} and {\it ROSAT}, another class of SNRs was identified with radio shells and center-filled X-rays; these sources are called mixed-morphology or thermal-composite SNRs \citep{rho98,shelton99}. 

\subsection{Effects of the Explosions on SNR Morphologies} \label{sec:explosions}

\subsubsection{Distinguishing Type Ia and Core-Collapse SNRs} \label{sec:typing}

While SNRs were classified based typically on their radio and X-ray morphologies as described above, the relationship of those morphologies to the originating explosions (Type Ia or core-collapse) was unclear. The reason stemmed in part from the challenge of constraining the SN type of SNRs. Generally, SNe are classified phenomenologically based on their optical spectra and light curves near maximum brightness \citep{minkowski41,filippenko97}. Given that SNe are typed based on their properties days after an explosion, an alternative means is necessary to discern the explosions responsible for SNRs, which are hundreds to tens of thousands of years old. 

The most reliable ways to classify SNR explosion types are the detection of a central neutron star (although chance alignments are possible: \citealt{kaspi98}) or of spectra from surrounding light echoes (e.g., \citealt{rest05,krause08,rest08}). Another means to constrain the progenitor star is to measure the number of OB stars (e.g., \citealt{maggi16}) or to investigate the stellar populations (e.g., \citealt{badenes09,jennings12}) in the vicinities of SNRs. 

However, the most widely used method to type SNR explosions is to measure metal abundances using X-ray spectroscopy and compare the values to those predicted by SN models. In particular, the oxygen-to-iron ratio (O/Fe) is often employed for this purpose, since Type Ia SNe produce an order of magnitude more Fe than CC SNe, whereas CC SNe yield more intermediate-mass elements (e.g., O, Ne, Mg) than Type Ia SNe \citep{woosley02}. Using all of the approaches described above, many dozens of Milky Way and Magellanic Cloud SNRs have been identified as from Type Ia or from CC explosions. 

\begin{figure}[t]
\begin{center}
\includegraphics[width=\textwidth]{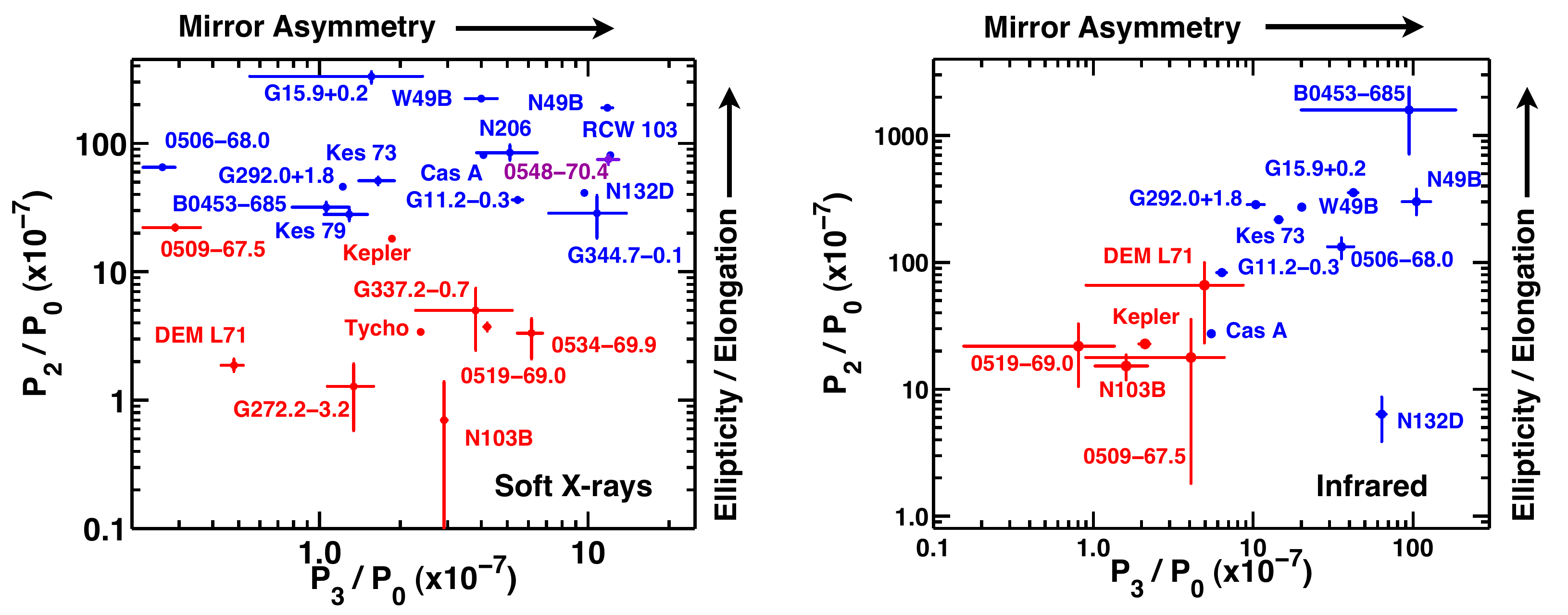} 
\end{center}
\caption{Results from calculation of the multipole moments (a power-ratio method) to 24 Galactic and LMC SNRs: quadrupole power ratio $P_{2}/P_{0}$ (which measures ellipticity/elongation) vs. octupole power ratio $P_{3}/P_{0}$ (which quantifies mirror asymmetry) of the {\it Chandra} X-ray (left) and of the {\it Spitzer} 24 $\mu$m (right) images. Type Ia SNRs are plotted in red points, and the CC SNRs are in blue (as classified by their abundance ratios, light echo spectra, neutron star detections, and/or their environments). The Type Ia SNRs separate naturally from the CC SNRs in both diagrams. Figures are adapted from \cite{lopez11} and \cite{peters13}.  \label{p2p3}}
\end{figure}

Using a sample of young SNRs with well-constrained explosion types from the literature, \cite{lopez09b} measured and compared the morphological asymmetries in the Si {\sc xiii} {\it Chandra} X-ray images of Type Ia and CC SNRs using a multipole expansion technique called the power-ratio method (PRM; see \citealt{jeltema05} and \citealt{lopez09a} for the mathematical formalism). These authors found that the Type Ia SNRs are more circular (have smaller quadrupole power ratios) and symmetric (have smaller octupole power ratios) than CC SNRs. In subsequent works, they extended the PRM to soft (0.5--2.1 keV) {\it Chandra} X-ray images and to {\it Spitzer} 24 $\mu$m images \citep{lopez11,peters13}, and the results were consistent in those bandpasses as well  (see Figure~\ref{p2p3}). These works suggest that the X-ray and infrared morphologies of SNRs may be used to constrain progenitors' explosions.

The SNR results described above are consistent with mounting evidence that Type Ia and CC SNe have intrinsically different symmetries. For example, spectropolarimetry studies -- the measure of the polarization of light as a function of wavelength as it is scattered through the debris layers of expanding SNe -- demonstrate that CC SNe have larger deviations from spherical symmetry than Type Ia SNe (e.g., \citealt{leonard06,wang08}. In some cases, the asymmetries of CC SNe can be substantial, altering the nucleosynthetic products of the explosions (e.g., \citealt{maeda03}), ejecting metals at higher velocities (e.g., \citealt{mazzali07}), and kicking newly-formed neutron stars to speeds $\ls$1000 km s$^{-1}$ \citep{lyne94,arz02,hobbs05,faucher06}.

It is worth noting that a systematic analysis of the Fe K-shell emission in Type Ia and CC SNRs recently showed that the Fe-K$\alpha$ centroid can also be used to type SNRs \citep{yamaguchi14}. Specifically, these authors found that Type Ia SNRs have lower energy Fe-K$\alpha$ centroids than those of CC SNRs, indication that the Type Ia SNRs have less ionized plasmas than CC SNRs. This result indicates that Type Ia progenitors do not modify significantly their environments, and they reside in lower-density media than CC SNRs. Thus, although typing SNRs was a major challenge to the community twenty years ago, researchers now have a variety of tools to diagnose the explosive origins of SNRs. 

\subsubsection{Connecting Neutron Star Kicks and SNR Morphologies} \label{sec:kicks}

Neutron stars are thought to possess a substantial `kick" velocity due to an asymmetric, CC explosion. Evidence for this is that NSs in the Milky Way are observed to have velocities of hundreds of km s$^{-1}$ \citep{lyne94,arz02,hobbs05,faucher06}, up to $\sim$ 1000 km s$^{-1}$ (e.g., \citealt{chatterjee05,winkler07}). Such large velocities are more than can be accounted for by the disruption of a close binary system (which produce velocities of $\sim$100 km s$^{-1}$: \citealt{lai01}), suggesting an asymmetry in the dynamics of the explosion is the likely origin of NS kicks.

Two theories have been proposed to explain NS kicks. One scenario is that hydrodynamical instabilities lead to asymmetric mass ejection, accelerating the NS in a direction opposite to the bulk of ejecta \citep{scheck06, wongwathanarat13,janka17}. Such models predict that the heavier elements (e.g., Fe, Ti) are expelled directly opposite to the NS kick, whereas the intermediate-mass elements (e.g., C, O, Ne, Mg) are only marginally affected. In another set of models,  a NS kick arises from anisotropic neutrino emission. In this case, the neutrinos carry the bulk of the explosion energy, and both the NS and the ejecta are expelled in the opposite direction \citep{fryer06}. 

\begin{figure}
   \includegraphics[width=0.49\textwidth]{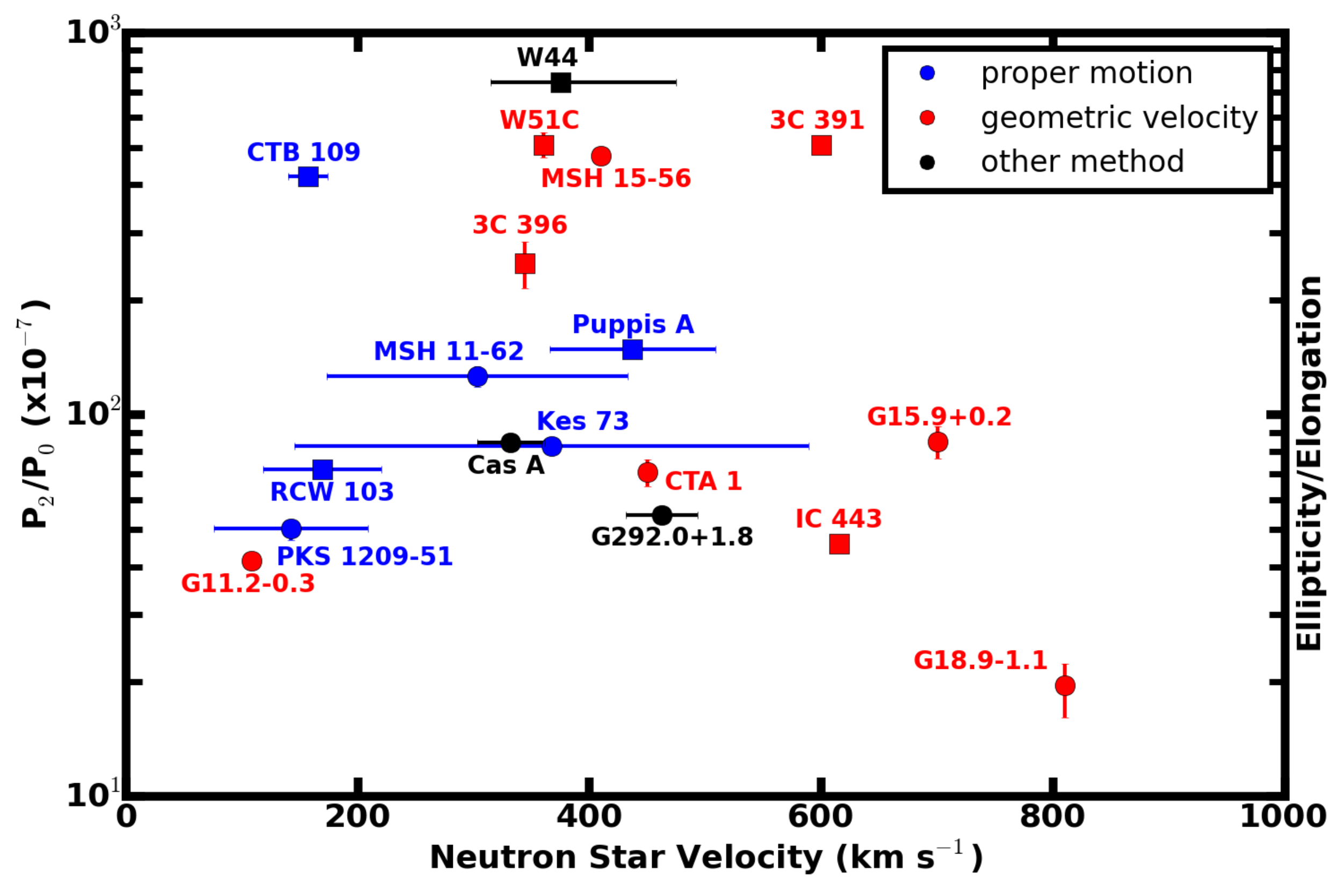}
 \includegraphics[width=0.49\textwidth]{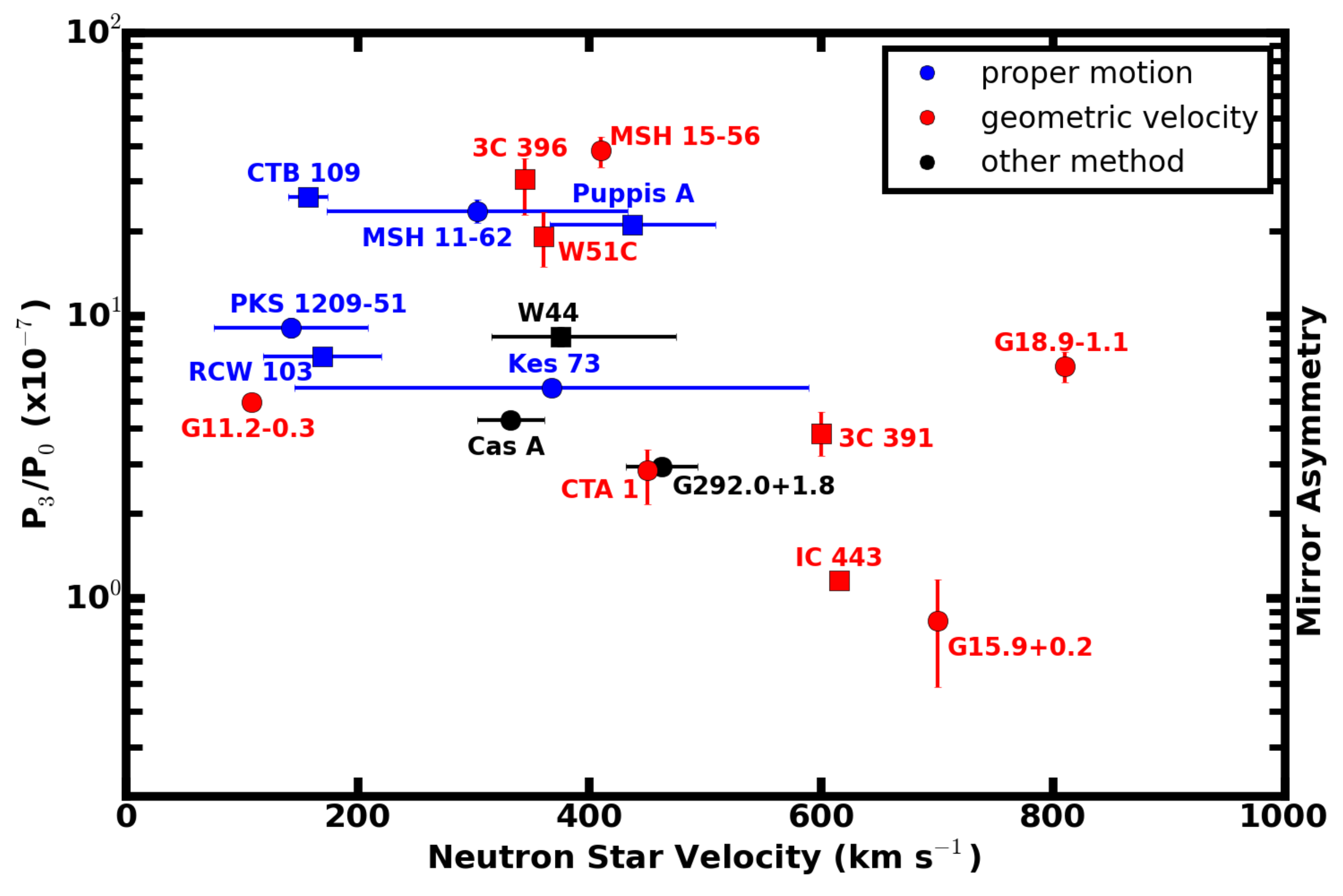}
\caption{The quadrupole (left) and octupole (right) power-ratios vs. neutron star velocities for a sample of 18 Galactic SNRs. Blue points are velocities measured via proper motions, and red points are velocities estimated from the spatial offset of the NS from the SNRs' geometric centers. Cas~A and G292.0$+$1.8's NS velocities are determined by back-evolved filament motion to find the explosion site; W44's NS velocity is determined by multiple methods. Circles are used for SNRs without evidence of CSM/ISM interaction, and squares are those with clear evidence of CSM/ISM interaction. Figure is adapted from \cite{hollandashford17}.}
\label{fig:NSkick1} 
\end{figure}

\begin{figure}
\includegraphics[width=\textwidth]{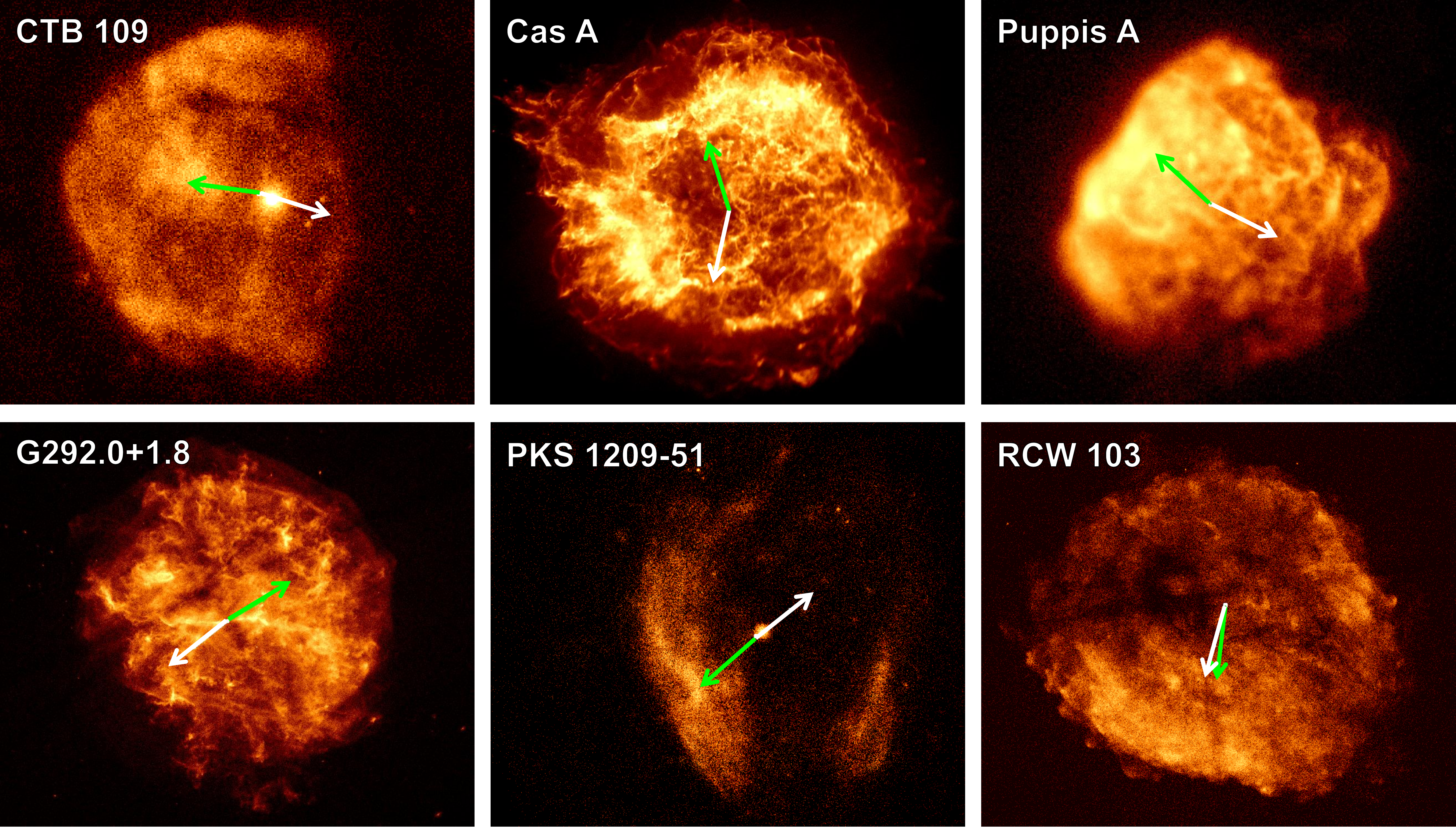}
\caption{0.5--2.1 keV {\it Chandra} and {\it ROSAT} images of the six Galactic SNRs with robust measures of their explosion sites that are analyzed by \cite{hollandashford17}. The green arrow points from the explosion site to the direction of the dipole moment; the white arrow represents the direction of the NS motion. In Cas~A and G292.0$+$1.8, the dipole moment direction reflects the bulk of ejecta emission. In CTB~109, Puppis~A, and RCW~103, the dipole moment points towards enhanced emission due to interactions with CSM/ISM or a molecular cloud. It is unclear if the emission comes from ejecta or interactions in PKS~1209$-$51. Figure is adapted from \cite{hollandashford17}.}
\label{fig:NS_dipole} 
\end{figure}

Recently, \cite{hollandashford17} investigated the relationship between NS kick velocities and their associated SNRs' X-ray morphologies of 18 sources observed by {\it Chandra} and {\it ROSAT}. These authors found no correlation between the magnitude of asymmetries (i.e., the quadrupole and octupole power ratios) and the NS speeds (see Figure~\ref{fig:NSkick1}). However, they also showed that the SNRs' X-ray emission (as measured via the dipole angle of their 0.5--2.1 keV images) was in the opposite direction as the NSs' motions in five out of six SNRs with well-constrained explosion sites (see Figure~\ref{fig:NS_dipole}). 

In addition, \cite{hollandashford17} also noted that the X-ray emission of two of the SNRs (Cas~A and G292.0$+$1.8) is dominated by the ejecta, and the other sources' morphologies are influenced at least in part by their complex environments. In a similar analysis to \cite{hollandashford17}, \cite{katsuda17} showed that silicon, sulfur, argon, and calcium are distributed opposite to NS directions of motion in six young SNRs. The results of both works are consistent with predictions from models where NSs are kicked by ejecta asymmetries \citep{scheck06,wongwathanarat13,janka17}.

\subsubsection{The Persistent, Elongated Morphologies of Jet-Driven SNRs} \label{sec:jets}

\begin{figure}[t]
\begin{center}
\includegraphics[width=\textwidth]{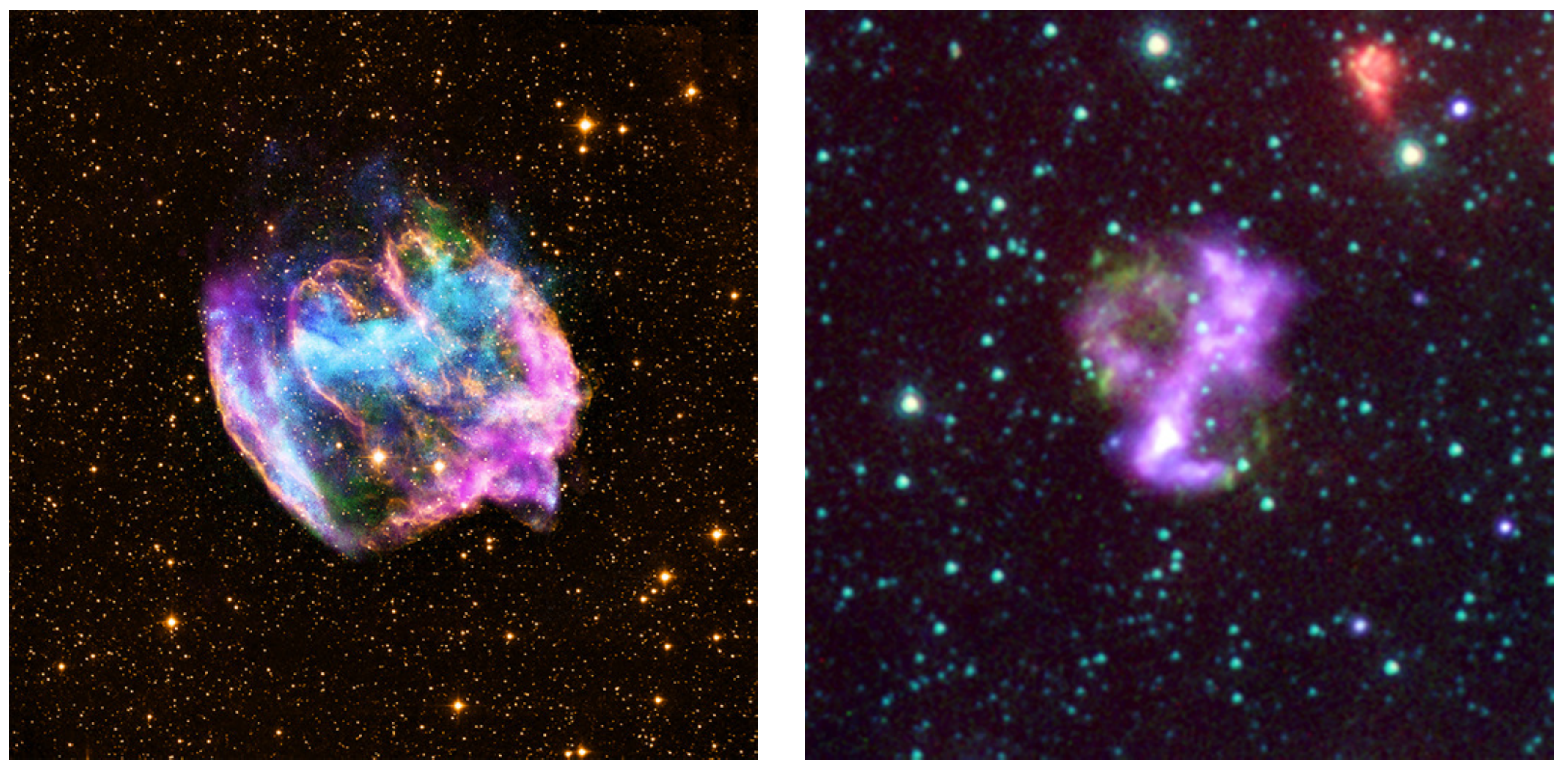}
\end{center}
\caption{Multiwavelength images of two extremely elliptical SNRs thought to be from jet-driven explosions (Lopez et al. 2013a, 2013b). {\it Left}: SNR W49B, with X-rays in blue, IR in yellow, and radio in purple. {\it Right}: SNR 0104$-$72.3, with X-rays in purple and IR in green. Images are from NASA/{\it Chandra} press releases. 
\label{bipolar}}
\end{figure}

The comparative, systematic analysis of SNR samples described in Sections~\ref{sec:typing} and ~\ref{sec:kicks} is also an important means to identify and examine outlier sources. For example, two SNRs from the X-ray morphological studies have anomalously large elongations compared to other young remnants: W49B in the Milky Way and SNR~0104$-$72.3 in the SMC. Both are extremely elliptical (see the X-ray images in Figure~\ref{bipolar}), and the spatial distribution of iron in W49B is remarkable distinct compared to the other elements detected in X-rays \citep{lopez09a}. Using spatially-resolved {\it Chandra} X-ray spectra, \cite{lopez13a} and \cite{lopez14} showed the mean abundance ratios in these SNRs are consistent with the predicted yields of jet-driven CC SNe. The identification of this kind of explosion in the SMC, where only 23 SNRs are known, may be evidence that jet-driven SNe occur more frequently in low-metallicity environments (since the SMC has a metallicity of $\sim$20\% solar). In fact, recent work on long-duration gamma-ray bursts (which are thought to arise from jet-driven explosions) prefer low-metallicity environments \citep{Mod08}. Thus, it is possible that these two SNRs are local analogues to long GRBs observed as cosmological distances. 

To test whether jet-driven explosions would retain their elliptical morphologies a few thousand of years after SN explosion, \cite{gonzalez14} performed two-dimensional simulations of a jet-driven SN, following its subsequent evolution as a SNR. Figure~\ref{fig:jetsim} shows the synthetic thermal emission maps derived from this work. These authors performed two separate hydrodynamical calculations, each equipped to describe the behavior of the bipolar outflow at two different epochs. They used a simulation with a general equation of state and a reaction network to determine the nucleosynthesis that accompanies a jet propagating through a massive star (left panel). As demonstrated in the abundance distributions in Figure~\ref{fig:jetsim}, the jetted outflow that is responsible for synthesizing $^{56}$Ni carries more energy and inertia than the products of incomplete silicon and oxygen burning. Thus, the distribution asymmetry in the abundance is preserved as the SN material sweeps up the external medium. 

\begin{figure}[t]
\begin{center}
\includegraphics[width=\textwidth]{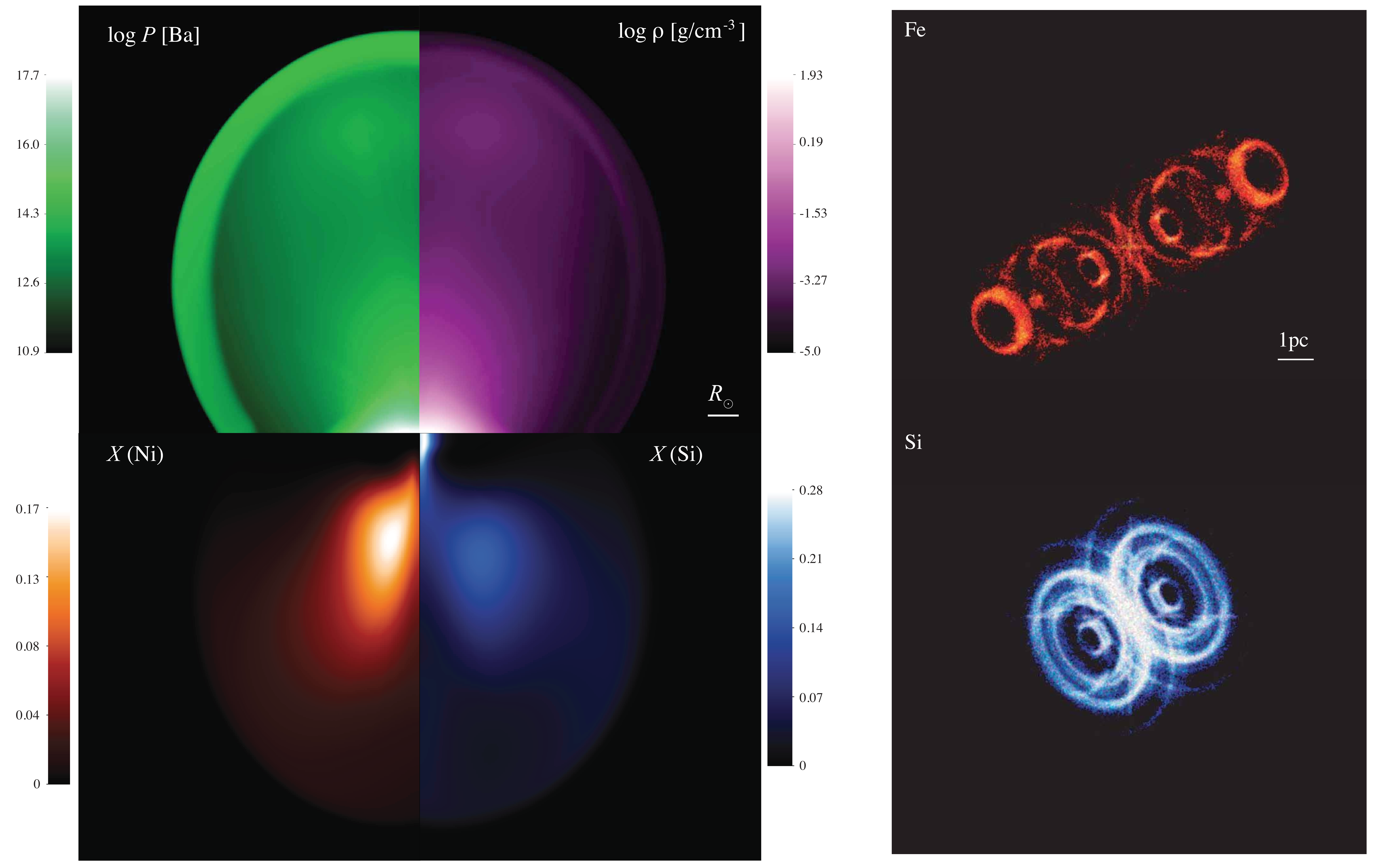}
\end{center}
   \caption{The X-ray properties of a jet-driven SN remnant as seen in simulations.  {\it Left Panel:} A jet-driven supernova remnant. {it Left panel}: Jet-driven SN explosion at the onset of the homologous phase, which is achieved $100$~seconds after the bipolar ejecta has broken free from the $R_\ast = 0.43 R_\odot$  stellar progenitor. The pressure $P$, mass  density $\rho$ and the silicon $X$(Si)  and nickel $X$(Ni)  mass fractions are indicated in each frame with corresponding size scales in units of $R_\odot$. This model is then employed to calculate  the evolution of a jet-driven SN remnant inside the wind bubble structure expected around a 25 $M_\odot$ massive star for  the case of an ISM pressure typical of the high-density molecular region surrounding the bipolar SNR. {\it Right Panel:}  Fe {\sc xxv} and Si {\sc xiii} X-ray emission maps. The maps are obtained integrating  the respective emission coefficients  along the line of sight. The images are computed at an evolutionary age  of 700~yr. To match the characteristics of W49B, the results of the simulation have been rotated 45$^\circ$ with respect to the plane of the sky (in the direction of the observer), and 120$^\circ$ with respect to the $z$-axis.}
\label{fig:jetsim}
\end{figure}

The resulting outputs (i.e., the density, velocity, and compositional structure of the ejecta) were employed as the initial conditions in the calculation of the expansion and associated thermal X-ray emission produced by the hot shocked gas. The Fe {\sc xxv} and Si {\sc xiii} maps of the jet-driven SNR inside a wind bubble of a 25 $M_{\rm sun}$ star are shown in Figure~\ref{fig:jetsim}, right panel. Even in the presence of an uniform external medium, the X-ray morphology is dominated by the bright iron jet, with explosive oxygen-burning products (such as silicon and sulfur) enclosing the Fe, as observed in W49B. 

\subsection{Influences of the Environment and the Surrounding Medium on SNR Morphologies} \label{sec:environments}

Since the progenitors of CC SNe have short main-sequence lives ($\sim$3--50~Myr), their explosions are expected to occur within or near the dense media from which the massive stars formed. Consequently, it is common for CC SNRs to show traits of interaction with an inhomogeneous or dense CSM. For example, roughly one quarter of all Galactic SNRs show evidence of interaction with molecular clouds, such as the coincidence of OH masers (indicating the presence of shocked molecular gas: e.g., \citealt{wardle02}).

A SNR-molecular cloud interaction will have a profound influence on the X-ray morphologies and spectra of SNRs. Large-scale density gradients can result in substantial deviations from spherical symmetry, and many SNRs interacting with molecular clouds have highly elliptical/elongated shapes compared to non-interacting CC SNRs \citep{lopez14b,hollandashford17}. A large number of interacting SNRs are centrally bright with thermal X-ray emission, whereas their radio morphologies are shell-like (as described at the beginning of Section~\ref{sec:morphologies}). Known as mixed-morphology (MM) SNRs, $\sim$40 of these sources have been identified in the Milky Way \citep{vink12}. Based on X-ray observations, MM SNRs can have enhanced metal abundances \citep{lazendic06} and/or isothermal plasmas across their interiors. 

In young SNRs, shocks generate ionizing plasmas that slowly reach collisional ionization equilibrium (CIE). However, X-ray observations with {\it ASCA} first revealed evidence that MM SNRs can have ``overionized" plasmas \citep{kawasaki02,kawasaki05}. In these cases, the electron temperature $kT_{\rm e}$ derived from the bremsstrahlung continuum is systematically lower than the effective ionization temperature $kT_{\rm z}$ given by the line ratios. Subsequent {\it Suzaku} observations of MM SNRs detected radiative recombination continuum (RRC) features in their spectra (e.g., \citealt{yamaguchi09,ozawa09,uchida12}), conclusive evidence of overionization.

In the collisional plasmas of SNRs, overionization is a signature of rapid electron cooling, and the physical origin of this cooling remains debated. One proposed scenario is thermal condution, where hot ejecta in a SNR interior cools by efficiently exchanging heat with exterior material (e.g., \citealt{cox99, shelton99}). Alternatively, the cooling may arise from adiabatic expansion, where the SN blast wave expands through dense CSM into a rarefied ISM (e.g., \citealt{itoh89,moriya12,shimizu12}).

To ascertain which cooling scenario is responsible for overionization, the localization of the overionized plasma is crucial. Toward this end, \cite{lopez13a} performed a spatially-resolved spectroscopic analysis to compare $kT_{\rm e}$ (derived by modeling the continuum) to $kT_{\rm z}$ (measured from the flux ratio of He-like to H-like lines). They found that the overionized plasma is concentrated in the west and has a gradient of increasing ionization from east to west. Given that the ejecta has is colliding with molecular material in the east and is expanding unimpeded in the west, the results suggest that adiabatic expansion is the dominant cooling mechanism of the hot plasma.

\begin{figure}[t]
\begin{center}
\includegraphics[width=\textwidth]{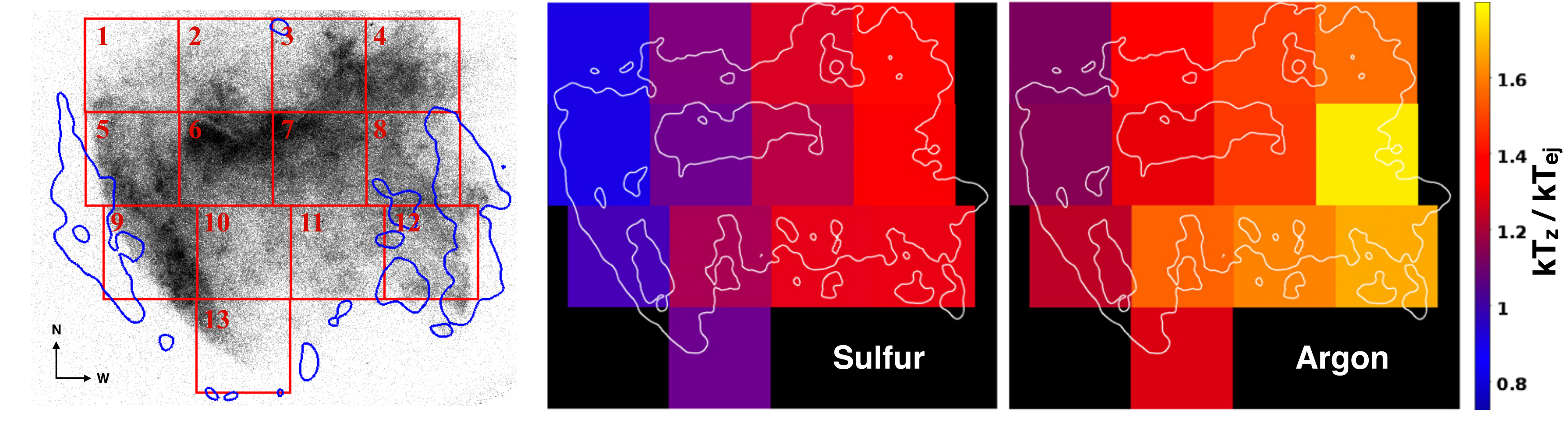}
\end{center}
   \caption{{\it Left}: 13 regions analyzed by \cite{lopez13b} to map overionization in the SNR W49B. Regions are overplotted on the full-band (0.5--8.0 keV) {\it Chandra} X-ray ACIS X-ray image, and the warm $H_{2}$ gas distribution is marked in blue The ejecta are colliding with the molecular gas in the east, whereas the ejecta are expanding freely in the west where the molecular material is thought to be in the foreground \citep{lacey01}. {\it Other panels}: maps of overionization across W49B. {\it Middle}: ratio of ionization temperature $kT_{\rm z}$ (from the ratio of S {\sc xv} to S {\sc xvi}) to the ejecta electron temperature $kT_{\rm ej}$ (from the bremsstrahlung continuum modeling). {\it Right}: maps of $kT_{\rm z}/kT_{\rm ej}$ using the ratio of Ar {\sc xvii} to Ar {\sc xviii}. Regions with $kT_{\rm z}/kT_{\rm ej} >1$ are overionized. Figures are adapted from \cite{lopez13b}.}
\label{fig:jetsim}
\end{figure}

\subsection{Instabilities \& Mixing} \label{sec:mixing}

On small scales, the distribution and scale of individual ejecta clumps can reveal the efficiency of the mixing between metals synthesized in the explosion and the surrounding CSM/ISM. Shock-heated ejecta expanding into lower-density material is subject to hydrodynamical instabilities (e.g., Rayleigh-Taylor instabilities) which are thought to be responsible for efficiently mixing ejecta metals (e.g., \citealt{blondin01,wang01}). X-ray and optical observations of individual SNRs show that these instabilities are capable of mixing ejecta out to and beyond the forward shock, in both Type Ia SNRs  (e.g., in Tycho: \citealt{hwang98}) and in CC SNRs (e.g., in Cassiopeia~A: \citealt{fesen96,hughes00,fesen11}).

\begin{figure}[t]
\begin{center}
\includegraphics[width=\textwidth]{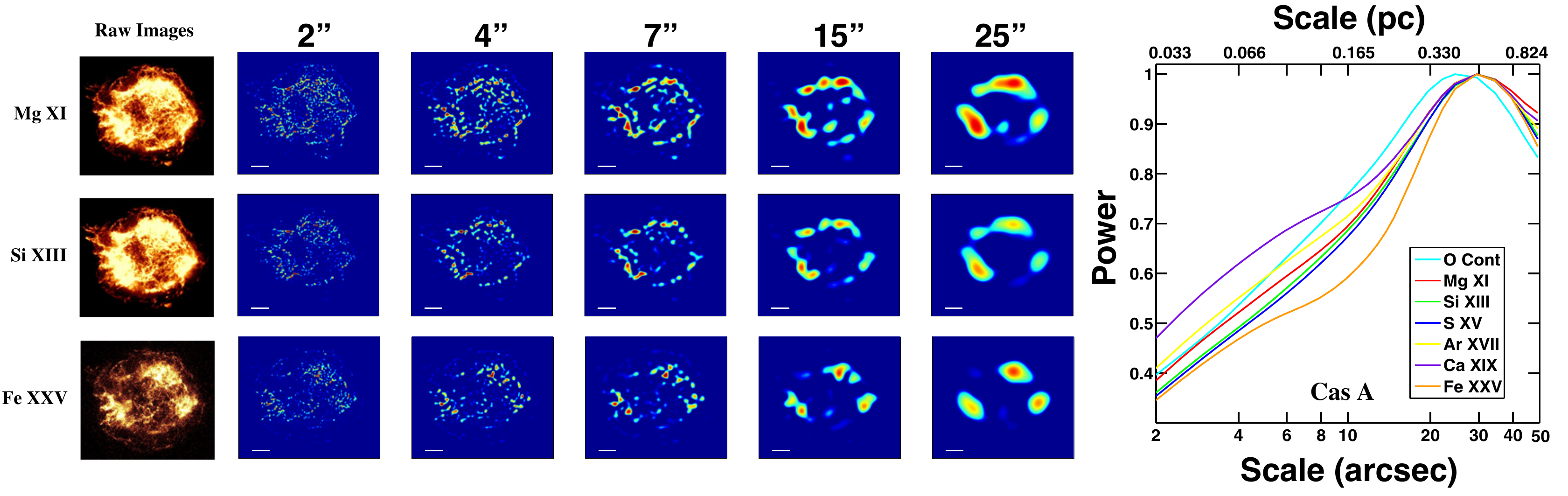}
\end{center}
   \caption{{\it Left panels}: Wavelet-transformed images of Cas~A showing the power at each scale for the Mg {\sc xi} (top row), Si {\sc xiii} (middle row), and Fe {\sc xxv} (bottom row) as observed by {\it Chandra}. {\it Right}: Plot of the relative power as a function of scale for narrow-band images corresponding to emission lines detected in the {\it Chandra} observations of Cas~A. The elements show similar profiles because they are well-mixed throughout the SNR, although some differences are evident (such as the Si {\sc xiii} emission in the northeast jet and the Fe {\sc xxv} located exterior to the Mg {\sc xi} and Si {\sc xiii} in the southeast). Figures are adapted from \cite{lopez11}.}
\label{fig:wta}
\end{figure}

\cite{lopez11} investigated the mixing of ejecta in nine SNRs using narrow-band {\it Chandra} images corresponding to emission lines from metals, including O, Ne, Mg, Si, and Fe. Specifically, they used wavelet-transform analysis to identify substructures and to measure their sizes. For example, Figure~\ref{fig:wta} displays the wavelet-transformed images of three emission line complexes (Mg {\sc xi}, Si {\sc xiii}, and Fe {\sc xxv}) in Cas~A. They found that all nine sources have well-mixed ejecta: 90\% of bright substructures of one element have corresponding substructures of the same size in another element, even if the metals arose from different burning processes. These findings reinforce observationally that hydrodynamical instabilities can efficiently mix ejecta in only a few hundred years.

\subsection{Bilateral SNRs and the Galactic Magnetic Field} \label{sec:bfield}

In the previous subsections, we have described how the thermal emission morphologies depend on the explosion and the environment. In addition, the morphology of the synchrotron radiation from SNR shocks gives clues about the surrounding medium, particularly the ambient density and the magnetic-field orientation. Specifically, \cite{gaensler98} examined a sample of bilateral SNRs at radio frequencies and showed that the bilateral axes tended to be aligned with the Galactic plane. However, some noteworthy exceptions existed, such as SN~1006, where the axis of symmetry is rotated almost 90$^{\circ}$ from the Galactic plane. 

\cite{orlando07} performed three-dimensional magnetohydrodynamical (MHD) simulations of a spherically-symmetric shock propagating through a magnetized ISM. They found that a gradient in ambient density or in magnetic-field strength can produce asymmetric bilateral SNRs, as in e.g., objects with two radio limbs of different brightnesses.

Recently, \cite{west16} investigated the radio morphologies of all Galactic axisymmetric SNRs as well, including those with only one bright limb. They found that using a simple model of SNRs expanding into an ambient Galactic magnetic field, they were able to reproduce the observed radio morphologies of the sample. One implication of this result is that if the large-scale B-field is known near a source, then the distance to that SNR can be inferred. In subsequent work, \cite{west17} demonstrated that the synchrotron radio morphologies of their bilateral sample are more consistent with quasi-perpendicular cosmic-ray (CR) acceleration than with quasi-parallel case.

\section{Kinematics of Young SNRs} \label{sec:kinematics}

As described in the previous section, the morphologies of SNRs are shaped in large part by their explosions but also by their environments. The other principal means to investigate the dynamics of the explosion mechanism and post-explosion expansion of SNRs is through kinematic studies of both radioactive and non-radioactive debris. In this regard, Cas~A is a fortuitous, nearby target. As the youngest Galactic core-collapse SNR known \citep{fesen06a}, it gives us a relatively high-resolution look at the expansion properties of the metals synthesized in the explosion.

Unless one views the Cas~A remnant as peculiar and not
representative of typical CC SN explosions, it can serve as a powerful and
unique case to investigate the kinetic properties of at least a subset and
possibly a significant fraction of CC events. In this Section, we highlight the
extensive work that has been done to characterize the kinematics of Cas~A and
what can be gleaned about the physics of explosions from this case study.

In terms of kinematics, Cas A is composed of three distinct sets of ejecta. The brightest at nearly all wavelengths is its thick main emission shell of ejecta. This material is shock-heated by the remnant's $\simeq$5000~km~s$^{-1}$ reverse shock and contains the bulk of Cas A's emitting debris. The shell's expansion is strongly asymmetrical with the rear hemisphere moving at roughly 6000 km~s$^{-1}$ while the front hemisphere is expanding around 2000 km~s$^{-1}$ slower.  Surrounding this shell of shocked ejecta lie hundreds of small ($\leq 0.5^{\prime\prime}$) ejecta knots with velocities between 7000 and 12,000 km~s$^{-1}$ with chemical compositions reflective of the outermost layers of the progenitor star at the time of explosion. 

\begin{figure}[t]
\begin{center}
\includegraphics[width=\textwidth]{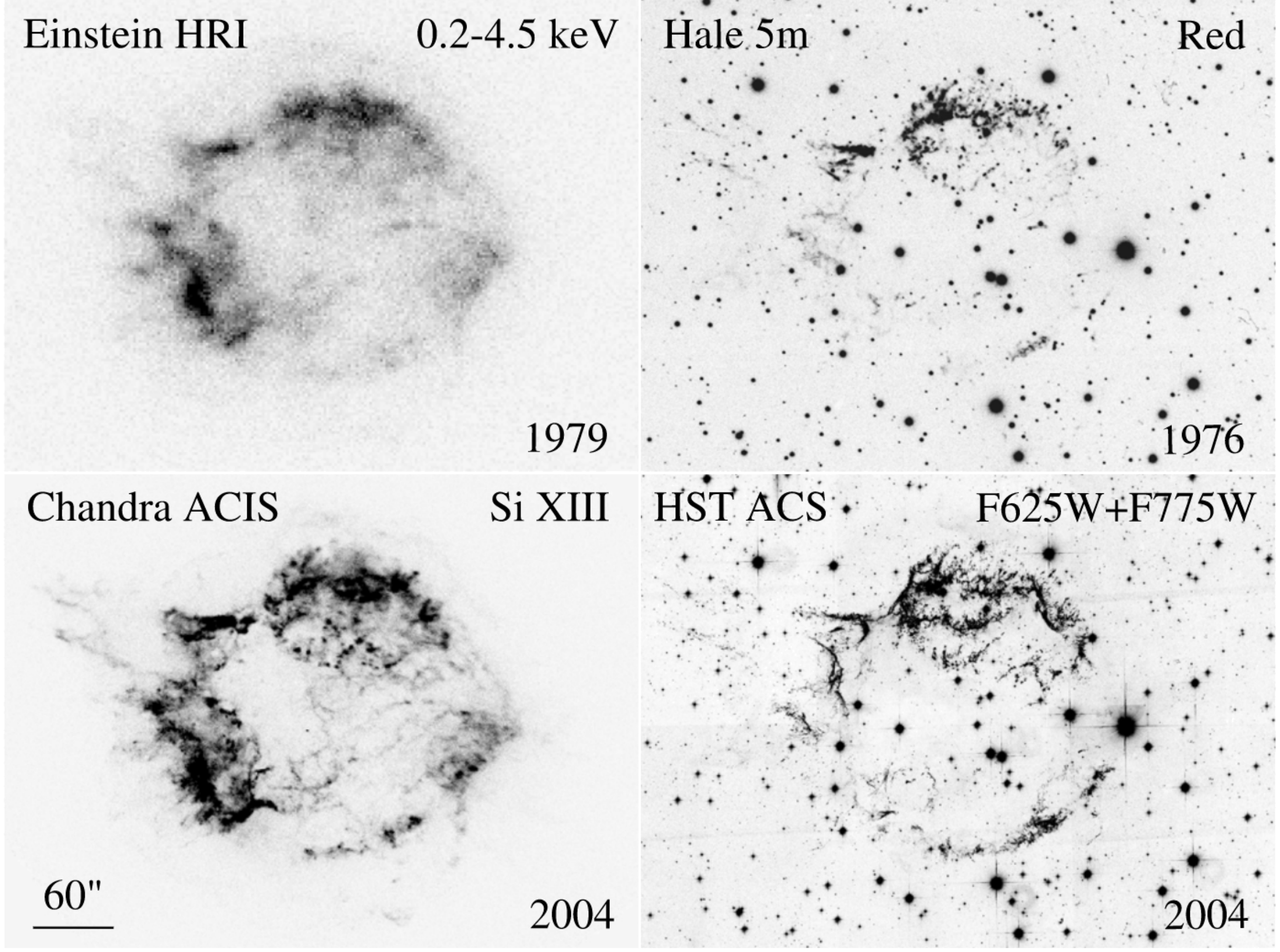}
\end{center}
\caption{
Comparison of emission differences of Cas A as seen in X-rays and
optical. Left hand panels show the 1979 {\it Einstein} X-ray image and a 2004 {\it Chandra}
continuum-subtracted Si {\sc xiii} X-ray image. Right hand panels show broadband red
optical images taken at similar epochs as the X-ray images. The optical images
are sensitive to line emission from [S II] $\lambda\lambda$6716,6731,
[O I] $\lambda\lambda$6300,6364, and [O II] $\lambda\lambda$7320,7330.
The Palomar Hale 5-m image is a 1976 July plate while the
{\it HST} image is a co-add of 2004 ACS F625W and F775W filter images \citep{patnaude14}.
  }
\label{fig:CasA_xray_n_opt}
\end{figure}

Cas A also exhibits two broad bipolar ``jets'' of especially high-velocity knots of ejecta (v = 8000--15,000 km s$^{-1}$) along its NE and SW limbs. These broad plumes of debris consist of ejecta streams containing hundreds of small, individual ejecta knots.  Interestingly, the fastest ejecta knots are richest in S, Ar, Ca  -- presumably from deep inside the progenitor -- whereas ejecta knots near the jets' base are O-rich, suggesting an explosive origin pushing interior material up and out through the outer layers. It seems likely that the energy for the creation of these ejecta plumes/jets is somehow associated with the central explosion engine.

{\it Chandra} and {\it Hubble Space Telescope} ({\it HST}) images of Cas A taken over the last
few decades have revealed a wealth of detail about its expanding main shell of
S-, O-, Ar-, Ca-rich ejecta.  However, it was obvious from the time of the first
high-resolution X-ray image of Cas A taken with the {\it Einstein} satellite in 1979 that
Cas A's X-ray and optical morphologies exhibit significant large-scale differences. 

This is shown in Figure~\ref{fig:CasA_xray_n_opt}
which contrasts the remnant's X-ray and optical emissions in the late 1970s and in 2004, 25 years later.
While there is some correlation between certain emission regions seen in the
1976 optical and 1979 X-ray images, especially along the remnant's northern
limb, there is a poor large-scale optical -- X-ray emission correlation over
much of the remnant, especially in the south, east, and west
quadrants.  A better correlation is seen between the X-ray and optical emission
in the 2004 {\it Chandra} and {\it HST} images.

Such comparisons make it clear that while Cas A's overall X-ray emission has
undergone only minor large-scale changes over the last few decades, its
optical appearance has gone from one of a faint and sparse emission structure to
a bright and extensive emission morphology \citep{patnaude14}.  Differences
between X-ray and optical images appear likely to be due to differences in the density of the
reverse-shocked material; that is, X-ray emission regions highlight low density,
interclump regions, whereas optical emission features are associated with much smaller, denser
debris knots.

\begin{figure}[h!]
\begin{center}
\includegraphics[width=\textwidth]{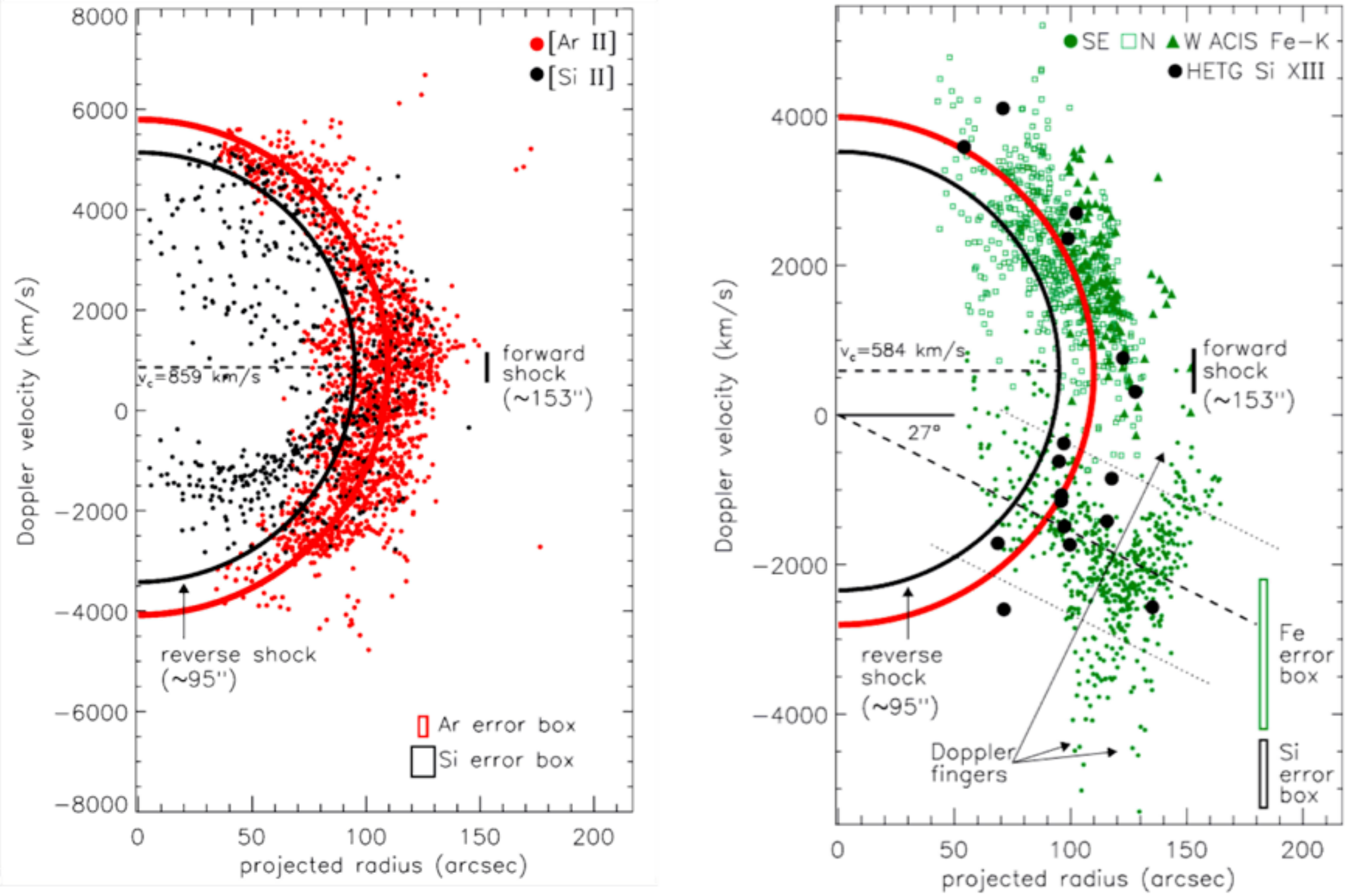}
\end{center}
\caption{Left panel: Doppler velocity vs. projected radius for infrared [Ar II]
emissions in red and [Si II] in gray. The mean velocity errors and spatial
resolution are indicated by error boxes at the bottom of the figure. The red
semicircle represents the best-fit spherical expansion  model.  The black
semicircle represents  the  reverse  shock  and  the projected radius of the
forward shock is also indicated. Right panel: Doppler velocity vs. projected
radius for X-ray Fe-K emissions in green and Si XIII emissions in black.
Different symbols are used to represent the north (open square), west (filled
triangle), and southeast (filled circle) Fe-K distributions \citep{delaney10}. }
\label{fig:delaney}
\end{figure}

Although the remnant's X-ray and optical emissions in Cas~A's main shell have
grossly similar morphologies in certain regions, they possess different expansion velocity ranges
and very different detailed structure.  Kinematic differences between
relatively cool and dense optical/infrared emitting knots and much hotter and
more diffuse X-ray emitting debris can be seen in Figure~\ref{fig:delaney}.
Such expansion velocity differences are discussed in more detail below in Section~\ref{sec:rings}. 

\begin{figure}[h!] 
\begin{center}
\includegraphics[width=\textwidth]{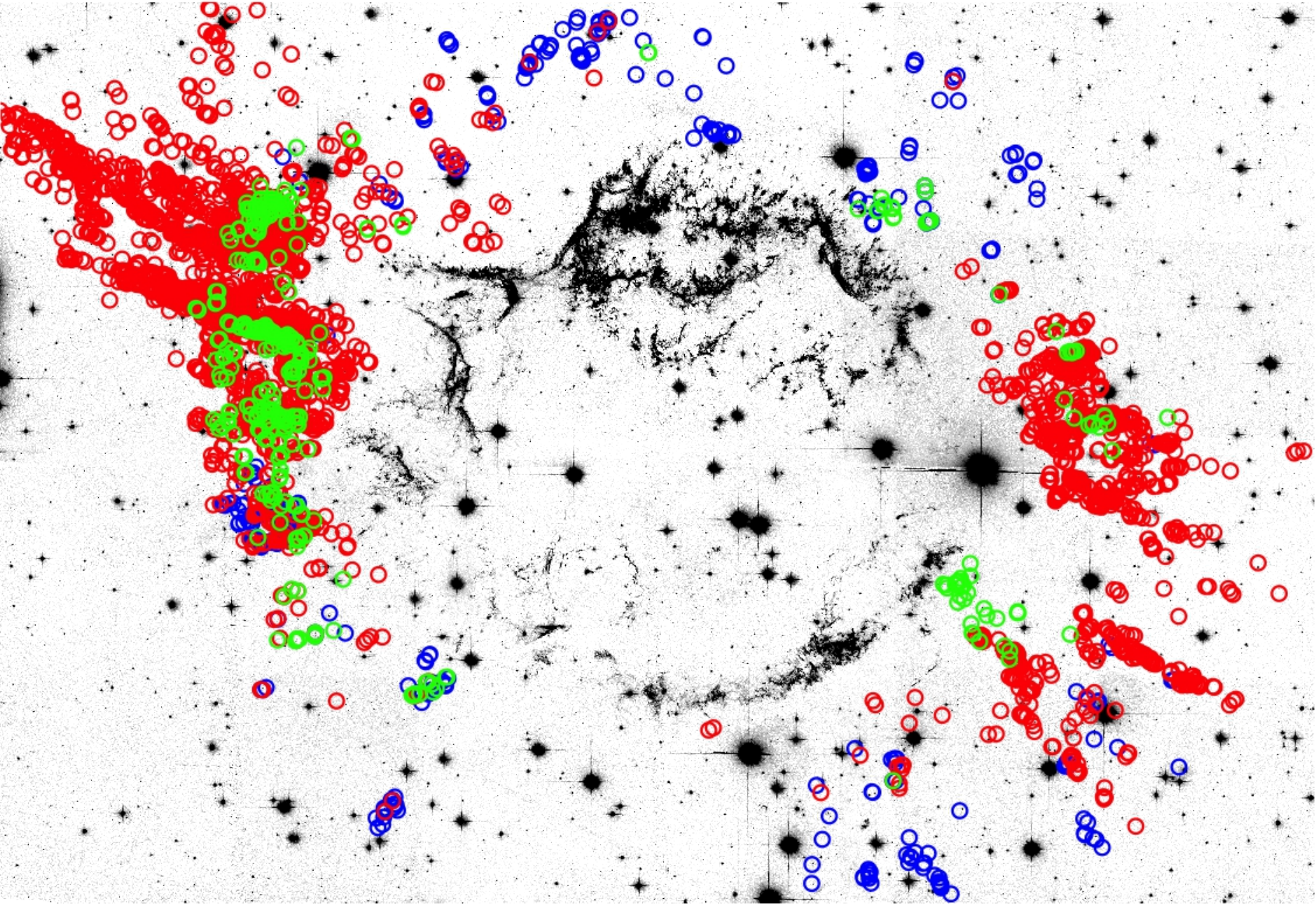} 
\end{center}
\caption{Locations of high-velocity ejecta outside of the remnant's main
shell of ejecta shown on a 2004 HST image.  Blue and green circles
indicate N and O emission knots while red circles mark S 
emission knots \citep{fesen16}. }
\label{fig:CasA_knots}
\end{figure}

Outside of the main shell of reverse-shocked material, {\it HST} images have
uncovered nearly 4,000 outlying fragments around the remnant and largely out
ahead of the 4500--5000 km s$^{-1}$ blast wave. These knots exhibit proper
motions indicating transverse velocities of up to 12,000 km~s$^{-1}$
\citep{fesen01,fesen06a}. They surround nearly all regions about the main shell
and are distinctly different in both chemistry and expansion velocity than the
remnant's NE and SW jets. Most of these knots are too small and faint to detect
from the ground. The expansion and chemical properties of these outer ejecta
knots is shown in Figure~\ref{fig:CasA_knots}.

As mentioned above, the presence of weak hydrogen emission lines in the spectra
of these outer knots (which indicate that the Cas~A progenitor had some
photospheric hydrogen at the time of outburst) is consistent with the SN IIb
classification based on light echo studies.  Faint streams of trailing emission
indicate ablated material off these outer SN fragments as they move through
the local ISM/CSM environment have been detected in deep {\it HST} images, showcasing
the earliest stage of ISM enrichment by SN debris. 

\subsection{Ejecta Rings and Interior Bubbles} \label{sec:rings}

\begin{figure}[h!]
\begin{center}
\includegraphics[width=\textwidth]{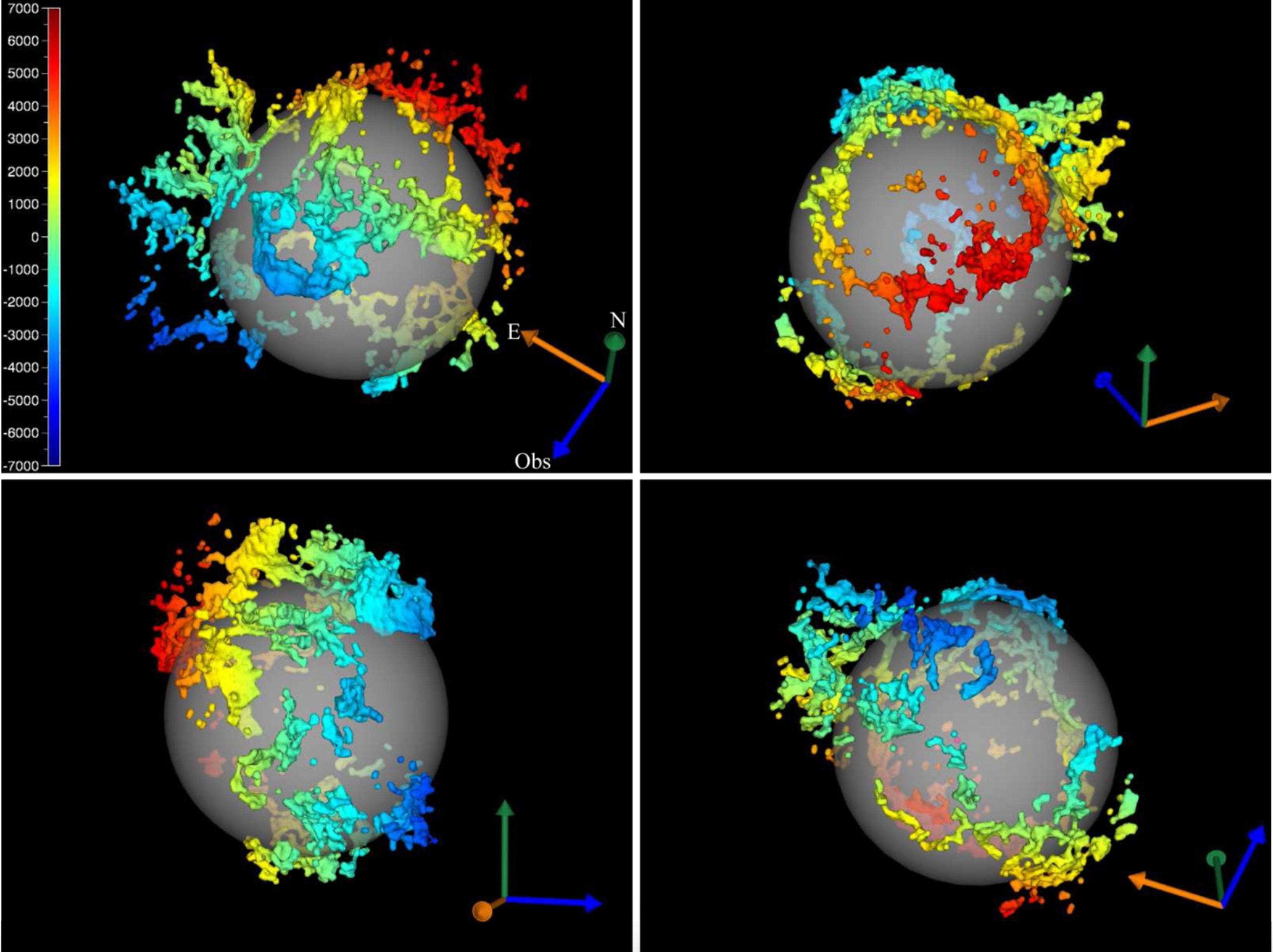}
\end{center}
\caption{Three-dimensional projections of the optical emission from Cas~A's main shell illustrating the
ubiquity of large ejecta rings.  Noticeable rings of ejecta include the blueshifted ring in the north and the much larger neighboring redshifted ring \citep{milisav13}.}
\label{fig:rings}
\end{figure}

\begin{figure}[t]
\begin{center}
\includegraphics[width=\textwidth]{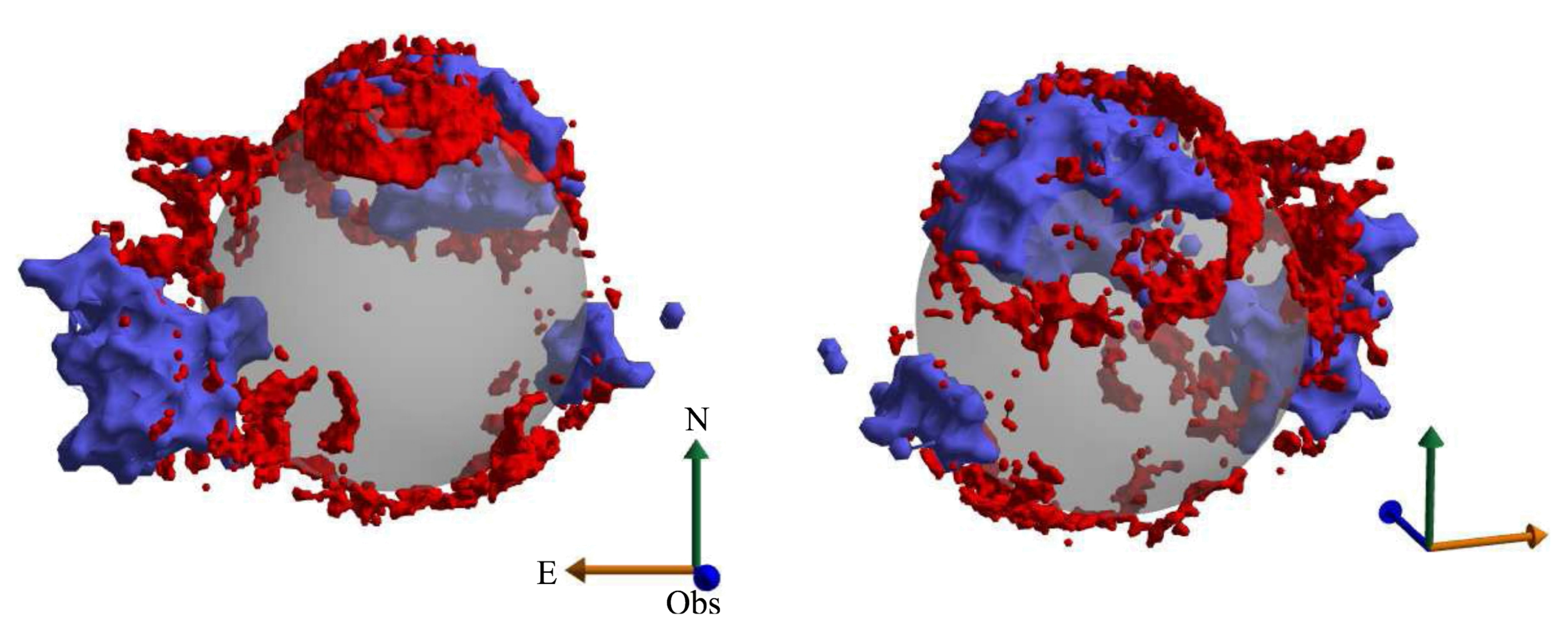}
\end{center}
\caption{Location of iron-rich X-ray emitting ejecta (blue) with respect to the
main shell sulfur- and oxygen-rich optically emitting ejecta (red). The X-ray
data shown are from DeLaney et al. (2010). 
Note in the right-hand panel the location of the iron-rich (blue) ejecta
well confined inside the large red ring of S,O-rich ejecta. Adapted from \citet{milisav13}.  }
\label{fig:DM_rings}
\end{figure}

Unlike the more or less random arrangement of debris expelled by a supernova
explosion one might have expected, the kinematic and chemical properties of Cas
A's main shell of ejecta has been found to consist of a series of large optical
and infrared rings of reverse shocked O-, S-, Ar-, Ca-rich ejecta. These rings
surround much of the remnant's X-ray emitting Fe-rich ejecta. This unexpected
chemical arrangement of ejecta can be seen in Figures~\ref{fig:rings} and
\ref{fig:DM_rings} where the remnant's main shell ejecta can be seen to lie
almost exclusively in ring-like structures.

\citet{reed95} and \citet{lawrence95} were the first to establish the
existence of conspicuous rings of reverse-shocked ejecta in Cas A.  More
recently, \citet{delaney10} and \citet{milisav13} identified about half a dozen
well-defined ring-like structures with angular diameters between 30$^{\prime\prime}$ and
120$^{\prime\prime}$ (0.5--2 pc).  Most of these rings are actually short cylinders giving
them a crown-like appearance in 3D projections. 

The height in velocity space of these crowns radially away from the
remnant's center is up to 1000 km s$^{-1}$.  This radial extent may be related to
the $\simeq$ 20--30 year radiative timescale for cooling of these reverse-shocked ejecta knots. 

The presence of these ejecta rings naturally led to questions about their
nature. \citet{delaney10} noted the coincidence of large ejecta
rings with the three regions of Fe-K X-ray emission and suggested that these
and other, less prominent features are regions where the ejecta have emerged
from the explosion as ``pistons" of faster-than-average ejecta. In this view,
the remnant's main shell rings represent the intersection points of these
pistons with the reverse shock, similar to the bow-shock structures.

An alternative explanation first suggested by \citet{blondin01} 
is that the observed ejecta rings
represent cross-sections of large cavities in the expanding ejecta created by a post-explosion input of energy from  plumes of radioactive
$^{56}$Ni-rich ejecta.  From {\it Spitzer} infrared and {\it Chandra} images and spectra,
\citet{delaney10} were able to show that Fe-rich X-ray emitting material was
surprisingly confined inside the remant's optical/IR emitting ejecta rings.  This structure
can be seen in Figure~\ref{fig:DM_rings} where the Fe-rich ejecta (blue) is found inside rings of S-, O-rich ejecta (red). 

Turbulent  motions  that  would  initiate  this  Ni  bubble structure  in  Cas~A  are  not  unlike  recent  3D simulations  of the  large-scale  mixing  that takes place  in  the  shock-heated stellar  layers  ejected  in  the explosion of a  15.5 $M_{\odot}$ blue supergiant star presented in \citet{hammer10}.  As shown in their models, a  progenitor's metal-rich core is partially turned over with nickel-dominated fingers overtaking oxygen-rich  bullets.  Although the evolution  of  these  simulations  is strongly dependent on the internal structure of the progenitor star \citep{ugliano12}, it is still tempting to draw an association between the Ni-rich outflows seen in the  \citet{hammer10} models and the rings of Cas A.

Discovery of Cas A's ejecta rings was followed-up by the detection of unshocked
material inside the remnant \citep{isensee10,milisav15}.  Cas A's interior
material consists of a handful of large cavities or bubbles, 
likely the result of  plumes of radioactive $^{56}$Ni which expand, pushing
non-radioactive material like O, S, Ar, and Ca into bubble-like structures. Such
an arrangement explains the observed Fe-rich ejecta inside rings of O-, S-rich debris. 
When the expanding bubbles encounter the remnant's reverse shock, the
intersection is in the form of rings filled with Fe-rich ejecta, the product of
$^{56}$Ni decay. 

\begin{figure}[h!]
\begin{center}
\includegraphics[width=0.95\textwidth]{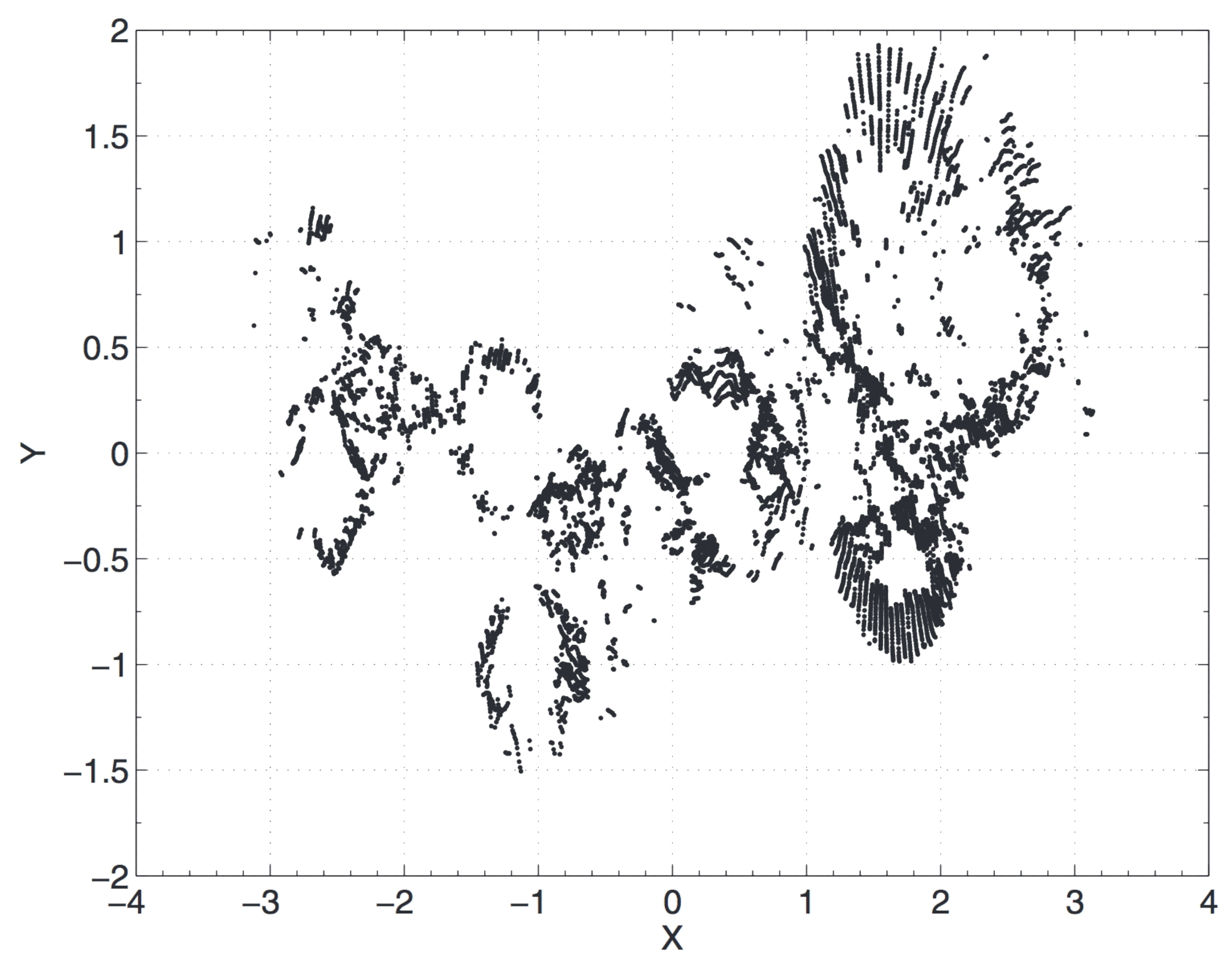}
\includegraphics[width=0.95\textwidth]{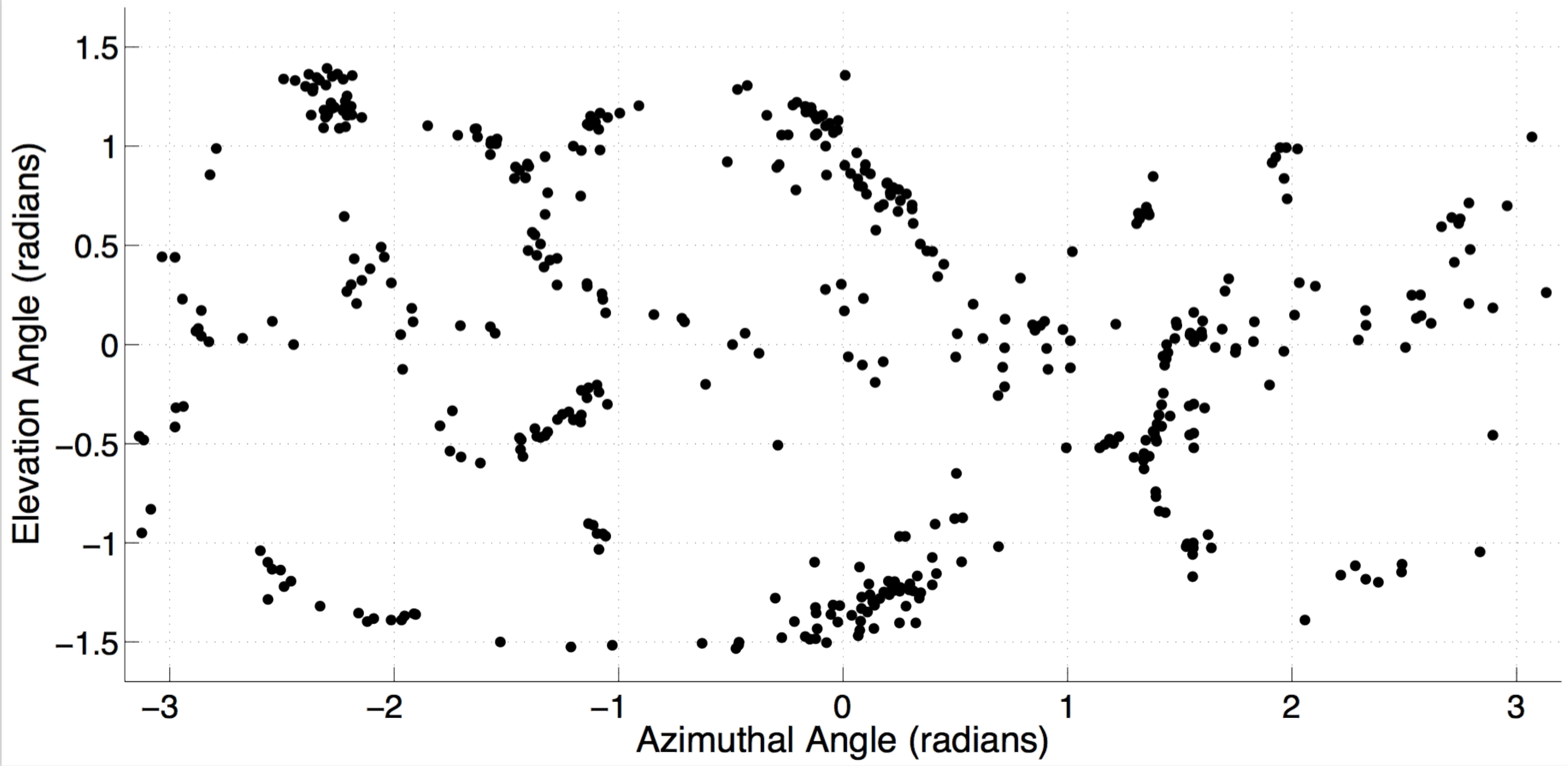}
\end{center}
\caption{ Mercator projections of ejecta in Cas~A and 3C~58.  Upper Panel: The
main shell of Cas~A's optically emitting ejecta \citep{milisav13}. Lower Panel:
High-velocity emission knots in the Crab-like SNR 3C~58 (Fesen et al. in prep.).
In both projections, the linear scale is equal in all directions around any
point and conformal, but the cylindrical map projection distorts the size and
shape of large objects, especially toward the poles.}
\label{fig:ejecta_maps}
\end{figure}

Consequently, it now appears that the rings of ejecta seen nearly
everywhere in Cas~A are a product of being the cross-section of large, internal
cavities when they encounter the inward moving reverse shock front. Similar
rings of ejecta to those seen in Cas A, thought to be due to an interior
bubble-like structure, are present in other CC SNRs.

Figure~\ref{fig:ejecta_maps} shows a comparison of Mercator maps of emitting
ejecta for Cas A (upper planel) and for the much lower expansion velocity and
Crab-like SNR 3C~58 (lower panel). In both cases, only a handful of rings -- not dozens --
are present in these two CCs SNRs despite the difference in their estimated
progenitor masses: 15--20 $M_{\odot}$ for Cas~A \citep{young06} and
8--10~$M_{\odot}$ for 3C~58. 

The presence of large-scale ejecta rings in CC SNe may help explain the clumpy
emission-line profiles seen in late-type spectra of SNe II/Ib events.
\citet{milisav12} showed how Cas~A's main shell spectra would appear as an
unresolved extragalactic source. Similarities were seen between the integrated
Cas~A spectrum and several late-time optical spectra of decades-old
extragalactic SNe.  Particularly well matched with Cas~A were SN 1979C, SN
1993J, SN 1980K, and the ultraluminous SNR in NGC~4449. 

Since the emission-line substructure observed in the oxygen forbidden line
profiles of Cas A is associated with the large-scale rings of ejecta,
\citet{milisav12} suggested that similar features in intermediate-aged SNe
(which have often been interpreted as ejecta ``clumps" or ``blobs") are, in
fact, probable signs that large-scale rings of ejecta are common in these
sources.

In summary, kinematic structures in CC SNe are highly dependent on the elemental abundances of the ejecta.  This is apparent in Cas A's main shell where Fe-rich ejecta is largely confined to certain regions surrounded by nonradioactive SN debris. It is also true regarding the higher-velocity ejecta knots around the remnant and out ahead of the forward shock.  These ejecta knots are the only ones which contain any hydrogen, a signature that the progenitor SN had a thin layer of hydrogen, leading to a SN II spectrum at early epochs.  Lastly, Cas A's very high-velocity ejecta jets display a strong radial `upside-down' chemistry unlike anywhere else in the remnant.

\subsection{High-Velocity Ejecta ``Jets''}  

It is of interest to consider the presence of jet-like ejecta streams seen in Cas~A in the context of recent work on supernovae that has established a continuum of explosion energies extending from broad-lined SNe Ic associated with gamma-ray bursts to more ordinary SNe Ibc \citep{soderberg10,chak15}. The existence of jets across such explosive events suggests that a wide variety of jet activity may potentially be occurring at energies that are hidden observationally.

The nature of the anomalously high velocities of ejecta in Cas A's NE and SW regions has long been a puzzle.  Asymmetric debris structures produced by circumstellar  interaction  have  been  theoretically  modeled by \citet{blondin96}, who showed that  a  jet-like  feature  of SN ejecta can be generated in the progenitor's equatorial plane from pole/equator density gradients in the local CSM. Thus, these high-velocity regions could be secondary features caused by  instability-powered  flows  from  an  equatorial torus,  where the  explosion  axis  is  loosely  defined  by  the  X-ray iron-rich regions  found  in  the SE and NW.

However, the very prominent rupture-like features in Cas~A's main shell near the
base of the NE jet is visible in X-rays, optical, IR, and  radio  images  and are
certainly suggestive of an explosive formation  process.  Although  perhaps
not  indicative  of  a  jet-induced explosion, there is nonetheless
substantial evidence that the NE and SW jets are associated with core-collapse
explosion dynamics in some way.

\begin{figure}[h!]
\begin{center}
\includegraphics[width=0.47\textwidth]{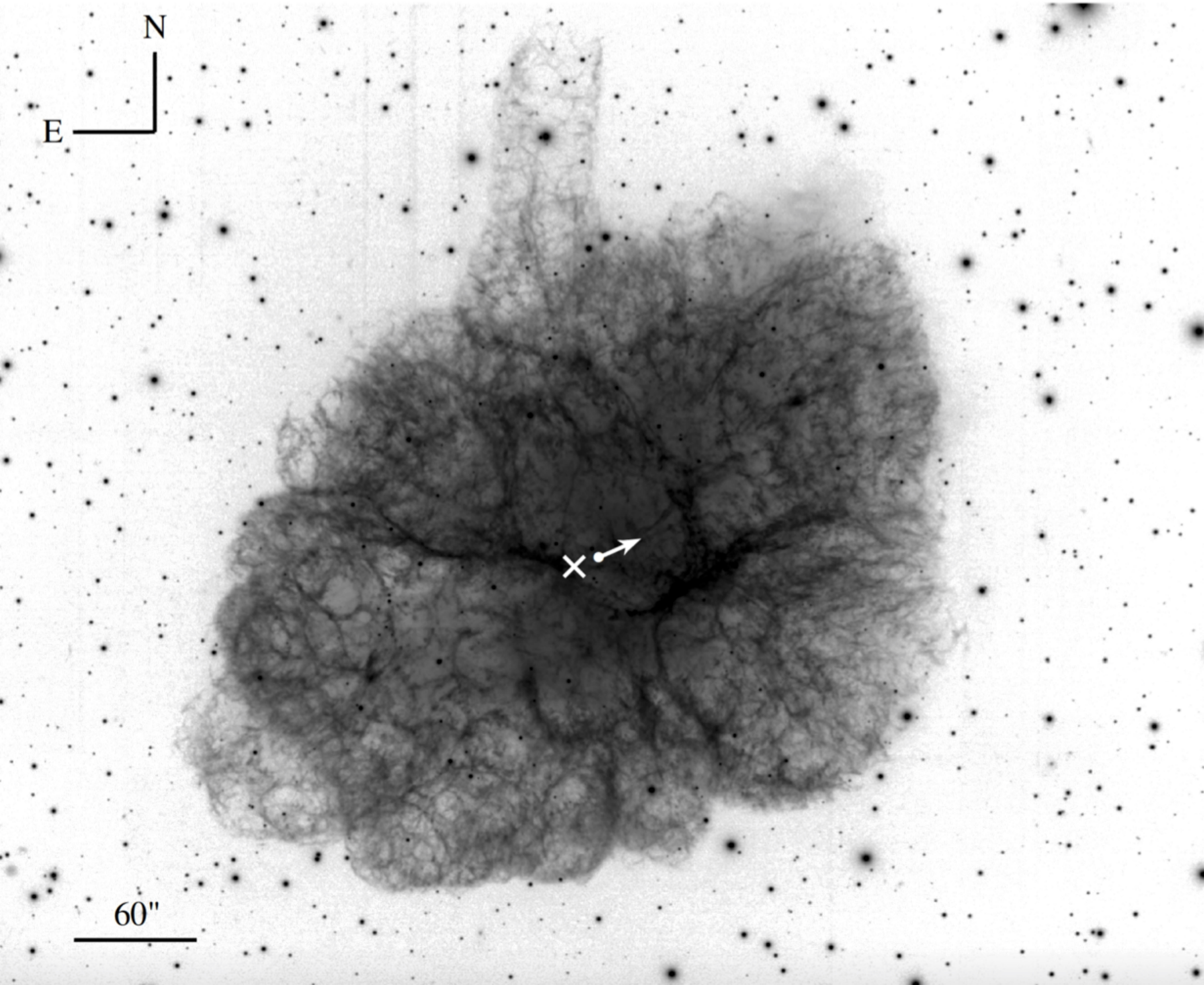}
\includegraphics[width=0.47\textwidth]{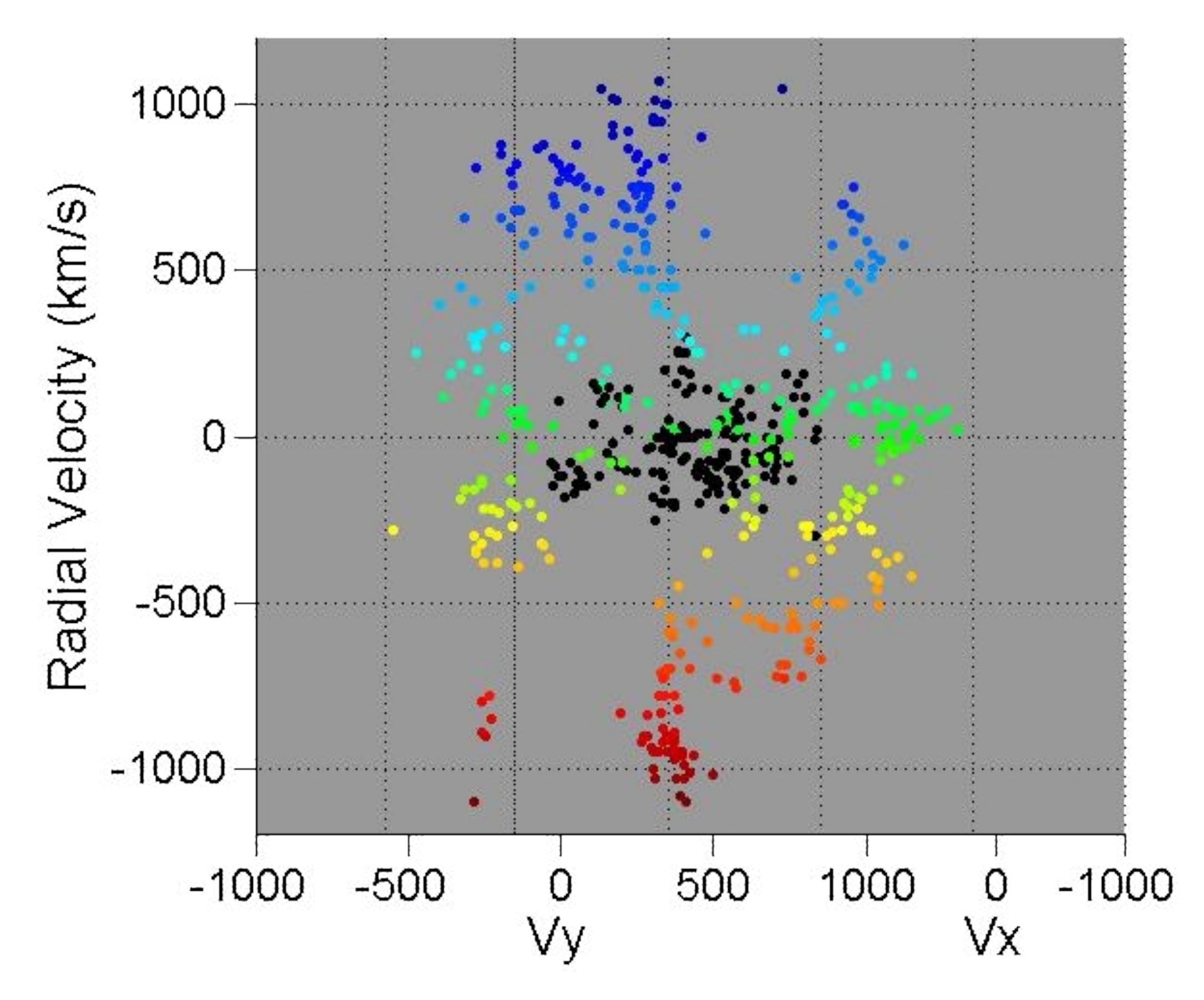}
\end{center}
\caption{Left Panel: An [O {\sc iii}] $\lambda\lambda$4959,5007 image of the Crab
Nebula showing its northern jet with respect to the entire remnant. The `X'
marks the \citet{nugent98} centre of expansion, and the white dot and arrow
marks the pulsar and its proper motion direction as determined by
\citet{kaplan08}.  Right Panel: 2D projection of observed radial
velocities (color-coded) vs.\ transverse RA and Dec velocities for optical emission knots in
the young Crab-like SNR 3C~58 assuming an explosion age of 2200 years. Slower moving CSM knots
are shown in black. Note that the wide north and south openings (jets) 
exhibit a similar broad opening morphology to that seen in the 
Crab's northern filamentary jet feature (Fesen et. al. in prep.). }
\label{fig:jets}
\end{figure}

The location of chemically distinct, optically-emitting knots in both  jets is also  consistent  with  an  unusual,  high-velocity ejection of underlying mantle material \citep{fesen01}. Knots exhibiting a mix of H$\alpha$, [N~{\sc ii}], [O~{\sc ii}], [S~{\sc ii}], and [Ar~{\sc iii}] line emissions are only observed in the two jet regions, consistent with an eruptive and turbulent mixing of the underlying  S-, O-, and  Ar-rich material  with photospheric H- and N-rich layers \citep{fesen96,fesen01}.  Furthermore, spectra of optical knots in the NE and SW jets lying farthest out and possessing the highest ejection velocities show no detectable emission lines other than those of [S~{\sc ii}] and [Ca~{\sc ii}], suggesting an origin from the Si-, S-, Ar-, Ca-rich layer deep inside the progenitor star.

Kinetic energy estimates for the Cas~A jets are around $1 \times 10^{50}$ erg \citep{fesen16}. An explosive origin of the jets would mean this value may be an underestimate since it does not include the energy needed to propel the material in both jets up through Cas~A's progenitor's outer layers. This added energy, together with the possibility of additional mass from undetected diffuse jet material, means the jets likely contain 10\% or more of the expected $10^{51}$~erg SN explosion energy.
 
Cas~A is not alone in being a young CC SNR with jet-like structures (see e.g., Section~\ref{sec:jets}). As shown in Figure~\ref{fig:jets}, low-mass progenitor CC SNe, such as the Crab Nebula and 3C~58, exhibit broad, plume-like ejecta streams. The fact that jet-like streams of unusually high-velocity material are present in both low-mass progenitors like the Crab and high-mass progenitors as in GRBs may be an indication that the central engine dynamics are not dis-similiar.

One additional clue as to the nature of Cas A's NE and SW ejecta jets and the explosion dynamics comes  from  examining  the  kinematic properties  of  the  inferred motion  of  the remnant's neutron star. Its location some 7$^{\prime}$ south of Cas A's expansion center (see Figure~\ref{fig:NS_dipole}), suggests a transverse velocity ``kick" of 350 km s$^{-1}$  imparted during the explosion \citep{fesen06b}. Intriguingly, the projected line connecting the NE and SW jets  lies  nearly perpendicular  to  the  inferred  direction  of  the NS (see Figures~\ref{fig:NS_dipole} and \ref{fig:CasA_knots}). Assuming the NE/SW axis to be the most significant in the original core-collapse, this runs counter to most jet-induced explosion models that predict that the neutron star will undergo a kick roughly aligned with the jet axis. 

\subsection{The Kinematics of Radioactive Titanium in Young SNRs}  

With the launch of {\it NuSTAR}, the first focusing, hard X-ray (3--79 keV) observatory \citep{harrison13}, it is possible to measure the kinematics of $^{44}$Ti from the radioactive decay lines at 68 and 78 keV. To date, these lines have been detected in two young CC SNRs: SN~1987A \citep{boggs15} and Cas~A \citep{gref14,gref17}. In the case of SN~1987A, the lines were narrow and redshifted with a velocity of $\sim$700 km~s$^{-1}$, and the lack of a blueshifted component suggests large-scale asymmetry of the ejecta.

\begin{figure}[h!]
\begin{center}
\includegraphics[width=\textwidth]{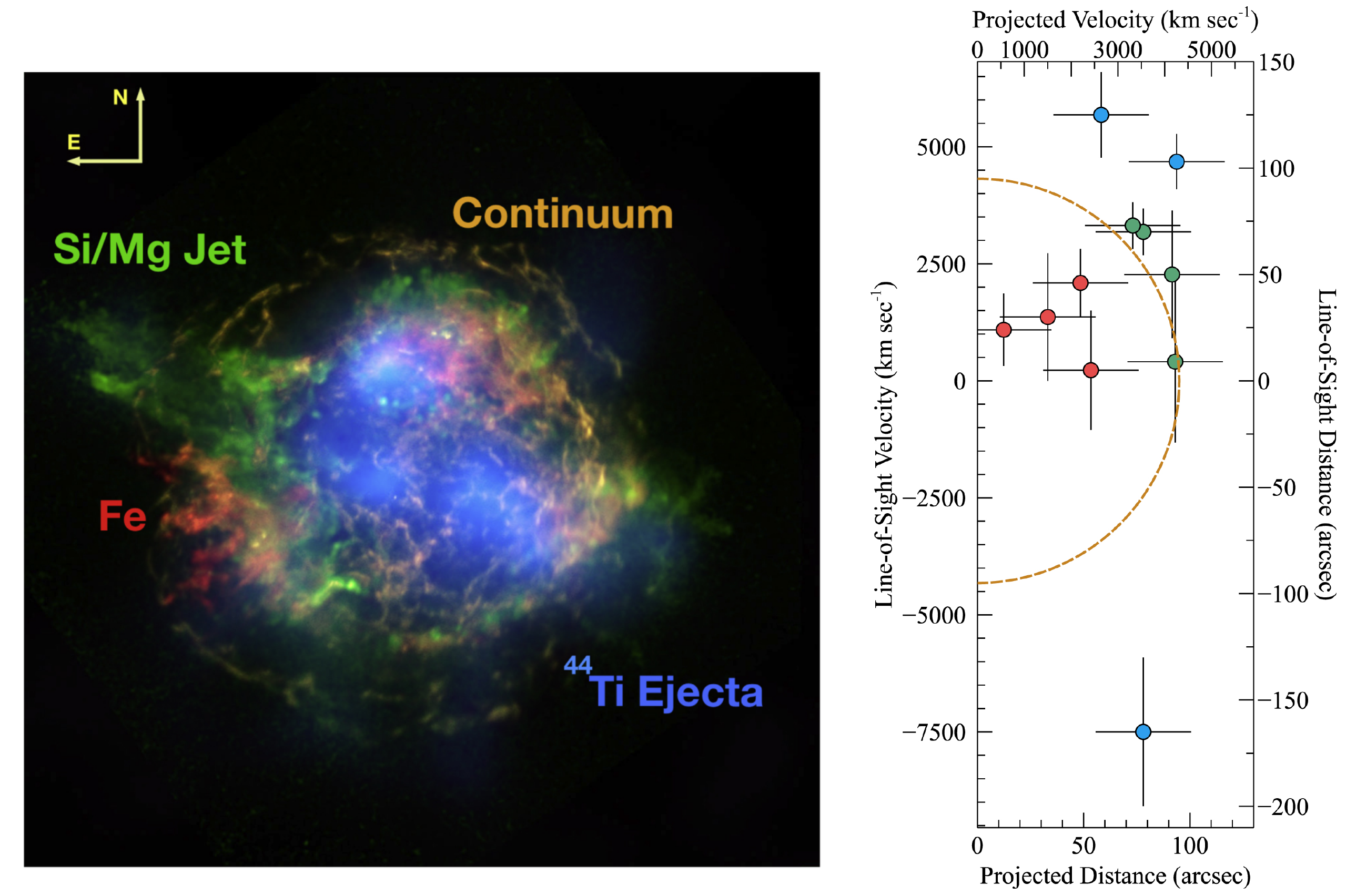}
\end{center}
\caption{Left: Three-color image of Cas~A, with $^{44}$Ti (from {\it NuSTAR} observations) in blue, 4--6 keV continuum in gold, Fe in red, and Si/Mg in green (the latter from {\it Chandra} observations). Right: The light-of-sight velocity of the detected $^{44}$Ti versus the projected distance from the center of expansion of Cas~A, where positive velocities indicate redshift. The gold dashed circle represents the location of the reverse shock; thus, the red points are interior to the RS, green points are at/near the RS, and blue points are exterior to the RS. Figures are from \cite{gref17}.}
\label{fig:nustar}
\end{figure}

In Cas~A, the $^{44}$Ti was localized using the {\it NuSTAR} observations, and one major surprise was that the $^{44}$Ti did not trace the distribution of Fe K (see Figure~\ref{fig:nustar}, left panel), nor does it appear associated with the NE/SW jets \citep{gref14}. Using deeper {\it NuSTAR} data, \cite{gref17} performed spatially-resolved spectroscopic analysis of the $^{44}$Ti ejecta and found that $\sim$40\% of the material is interior to the reverse shock, 40\% is at/near the reverse shock, and 20\% is beyond the reverse shock (see Figure~\ref{fig:nustar}, right panel). Generally, the $^{44}$Ti interior to the RS is not correlated with observed features in the optical or infrared. The $^{44}$Ti exterior to the RS is co-located with the shock-heated Fe, whereas regions with Fe do not necessarily have associated $^{44}$Ti. The authors interpreted this result as evidence that local conditions during the explosive nucleosynthesis varied enough to suppress the production of $^{44}$Ti in some regions. Based on the $^{44}$Ti line centroids, the majority of the material is receding along the line of sight, with velocities ranging from 1,100--5,700 km s$^{-1}$ (though there is some blueshifted emission as well). The flux-weighted average of the $^{44}$Ti velocities gives the direction of this ejecta component that is almost exactly opposite to the direction of the Cas~A NS, similar to the results presented in Figure~\ref{fig:NS_dipole}.

\section{Conclusions and Future Prospects}

The morphology and kinematics of SNRs are an especially powerful means to test and constrain SN explosion models. The predictive power of simulations has improved dramatically over the last few years, and the results can be compared to observational characteristics of SNRs, including large-scale compositional and morphological asymmetries as well as the three-dimensional kinematics of ejecta material. In particular, as the youngest known CC SNR in the Milky Way, Cas~A offers an up-close view of the complexity of these explosive events that are unresolved in extragalactic sources.

In the future, an exciting prospect in the study of SNRs is X-ray micro-calorimeters. CCD energy resolution is insufficient to resolve He-like and H-like line complexes of ions at X-ray wavelengths, and grating spectrometers are only useful when objects have minimal angular extension (e.g., SN~1987A: \citealt{dewey08}; SNR 1E~0102.2$-$7219: \citealt{flanagan04}) or if the ejecta knots are sufficiently bright and isolated (e.g., Cas~A: \citealt{lazendic06}; G292.0$+$1.8: \citealt{bhal15}). X-ray microcalorimeters are non-dispersive spectrometers that can take spatially resolved, high-resolution (with a few eV resolution) spectra across extended objects, like SNRs and galaxy clusters.

Before its untimely demise, {\it Hitomi} observed the LMC SNR N132D for only 3.7~ks \citep{hitomi17}, and it readily detected line complexes of S, Ar, and Fe. The Fe emission was highly redshifted at $\sim$800~km~s$^{-1}$, but no blueshifted component was detected, suggesting that the Fe-rich ejecta was ejected asymmetrically. In the shorter term (with an anticipated launch date of 2021), the replacement for {\it Hitomi}, the X-ray Astronomy Recovery Mission (XARM\footnotetext{See https://heasarc.gsfc.nasa.gov/docs/xarm/}), offers the chance of 5--7 eV energy resolution over a field-of-view of $\sim$3$^{\prime}$. 

Looking forward to the 2030s, the proposed {\it Lynx} X-ray Observatory is intended to have a microcalorimeter with high spatial resolution (1.0$^{\prime \prime}$ pixels) over a 5$^{\prime}$ field with a few eV energy resolution. These capabilities will enable three-dimensional mapping of the hot metals synthesized in SN explosions, similar to the results presented in Section~\ref{sec:kinematics} but at X-ray wavelengths. Additionally, with $\sim$50 times the effective area of {\it Chandra}, {\it Lynx} will detect fainter and more distant sources, such as SNRs in Local Group galaxies. In this way {\it Lynx} will dramatically increase the sample size of young SNRs with morphological and kinematic measurements. These observations will be crucial to inform SN models and probe SN feedback and enrichment in different Galactic environments.

\begin{acknowledgements}

LAL gratefully acknowledges that this work was supported through NSF Astronomy \& Astrophysics Grant AST$-$1517021. The authors thank the organizers of the workshop held in ISSI Bern on 3--7 October 2016 where many helpful discussions took place. 

\end{acknowledgements}

\bibliographystyle{spbasic}
\bibliography{SNR}




\end{document}